\documentclass[aps,pre,twocolumn,longbibliography,showpacs,amsmath,amsmath,amssymb,amsfonts,superscriptaddress]{revtex4-1}
\usepackage{graphicx}
\usepackage{dcolumn}
\usepackage{bm}
\usepackage{color}
\usepackage{multirow}
\usepackage{hyperref}
\usepackage{natbib}
\usepackage{xcolor}
\definecolor{myOrange}{rgb}{1,0.5,0}
\definecolor{myRed}{rgb}{0.8, 0.2, 0}
\definecolor{mygreen}{rgb}{0, 0.7, 0}

\usepackage{amsmath, amssymb}
\usepackage{bbold}

\begin{document}
\title{Dynamical phase transition to localized states in the two-dimensional random walk conditioned on partial currents}

\author{ Ricardo Guti\'errez}
\affiliation{Complex Systems Interdisciplinary Group (GISC), Department of Mathematics, Universidad Carlos III de Madrid, 28911 Legan{\'e}s, Madrid, Spain}
\author{Carlos P\'erez-Espigares}
\affiliation{Departamento de Electromagnetismo y F\'isica de la Materia, Universidad de Granada, Granada 18071, Spain}
\affiliation{Institute Carlos I for Theoretical and Computational Physics, Universidad de Granada, Granada 18071, Spain}

\begin{abstract}
The study of dynamical large deviations allows for a characterization of stationary states of lattice gas models out of equilibrium conditioned on averages of dynamical observables. The application of this framework to the two-dimensional random walk conditioned on partial currents reveals the existence of a dynamical phase transition between delocalized band dynamics and localized vortex dynamics. We present a numerical microscopic characterization of the phases involved, and provide analytical insight based on the macroscopic fluctuation theory. A spectral analysis of the microscopic generator shows that the continuous phase transition is accompanied by spontaneous $\mathbb{Z}_2$-symmetry breaking whereby the stationary solution loses the reflection symmetry of the generator. Dynamical phase transitions similar to this one, which do not rely on exclusion effects or interactions, are likely to be observed in more complex non-equilibrium physics models.
\end{abstract}


\maketitle

\noindent {\it Introduction---} The study of large fluctuations of dynamical observables
in lattice gas models has greatly improved our understanding of non-equilibrium statistical mechanics, as it has shed light on rare events in transport phenomena \cite{bodineau04a,bodineau06a,derrida07a}, has unveiled fluctuation relations of wide applicability \cite{evans93a,gallavotti95a,kurchan98a,lebowitz99a,maes99a,harris07a}, and has clarified the origin of many-body effects in classical \cite{garrahan07a,garrahan09a} and quantum systems \cite{garrahan10a,carollo17a,carollo18b}. These microscopic analyses have been complemented with approaches based on the macroscopic fluctuation theory, where similar phenomena are studied at a coarse-grained level \cite{bertini15a}. Both frameworks rely on statistical ensembles of trajectories \cite{touchette09a,lecomte07c}, for which large-deviation functions play an analogous role to that of thermodynamic potentials in equilibrium statistical mechanics \cite{callen1985}. 

Dynamical phase transitions (DPTs),  characterized by sudden changes in the structure of trajectories that are reflected in the statistics of dynamical observables, are among the most interesting phenomena that have been unveiled by these approaches. DPTs manifest themselves as non-analyticities in the large deviation functions, and occur not only in systems driven far from equilibrium, but also in equilibrium settings, even in situations where the statistics of (static) configurations is trivial \cite{garrahan09a}. In recent years, the study of DPTs has been enriched by the application of the generalized Doob transform, which provides the subset of trajectories sustaining a given rare fluctuation \cite{simon2009,popkov10a,jack2010,chetrite13a,chetrite15b}, allowing for the characterization of dynamical phases.

The study of DPTs is frequently accomplished in one-dimensional systems \cite{bertini05a,bodineau05a,bertini06a,hurtado11a,perez-espigares13a,jack15a,tsobgni16a,baek17a,harris17a,lazarescu17a,perez-espigares18a,perez-espigares18b,perez-espigares19a,banuls19a,hurtado-guti20a}, where the computation of large deviations for large systems is analytically tractable. Analyses of higher-dimensional settings are challenging, since they rely on variational procedures and diagonalization problems whose complexity increases with the spatial dimension. Yet the statistics of total currents (quantifying the net number of particles crossing a section that spans the whole system) or  dynamical activities (counting configuration changes in a trajectory) have been studied in two dimensions, where a myriad of phenomena, including DPTs, have been reported \cite{hurtado11b, villavicencio14a, perez-espigares16a, tizon-escamilla17b, jack20a,casert2020}. In the case of a two-dimensional random walk on a lattice, the Doob-transformed dynamics that sustains a given fluctuation of the total current amounts to the inclusion of a uniform driving field. This implies that large fluctuations are achieved by flat, i.e.~lacking spatial structure, density and current fields, giving rise to current distributions that are trivially Gaussian. The fluctuations of so-called partial currents, i.e.~currents of particles that move across some `wall' or `slit' that does not span the whole system, are strikingly different, however, as we shall show.

In this paper we consider the statistics of partial currents across a finite slit of random walks on a square lattice, as well as its continuum hydrodynamic limit. We find that, despite the simplicity of the model, there is a DPT underpinning rare fluctuations of such partial currents, which is continuous and characterized by a spontaneous breaking of the $\mathbb{Z}_2$ reflection symmetry in the direction of the slit.
While small fluctuations are characterized by the formation of bands, 
rare fluctuations condense into localized vortices. 
This highlights how large fluctuations of partial currents, which are more relevant than total currents in experimental settings, are created by non-trivial emerging structures associated with a DPT.
Given the simplicity and generality of the model, similar DPTs are likely to be found in other systems with exclusion effects and interactions, and might shed light on intriguing results previously reported for the two-dimensional simple exclusion process \cite{bodineau08a}. Regarding the observability of the DPT under consideration, it seems realistic that it may be observed in experiments such as those devised to study global current fluctuations in diffusive systems \cite{kumar15a,falasco16a}. In fact, experiments designed to measure the statistics of other local observables have been successfully carried out in different settings, see e.g. Refs.\cite{ciliberto98a,feitosa04a}.

\noindent {\it Model---} We consider $N$ random walkers on a two-dimensional square lattice comprising $L\times L$ sites with periodic boundary conditions in the presence of a driving field ${\bf E} = (E_x,E_y)$, which induces asymmetric jump rates between neighboring sites.  The stochastic dynamics of each particle is governed by the master equation $\partial_t p(n,t) = - p(n,t) + \sum_{m\in \text{nn}_n} \Gamma_{n\leftarrow m}\, p(m,t)$,
where $p(n,t)$ is the probability of occupation of site $n\in \{1,2,\ldots,L^2\}$ at time $t$, and $\text{nn}_n$ contains its (four) nearest neighbors.
Written in operatorial form, $d{\bf p}(t)/dt = \mathbb{W} {\bf p}(t)$, with the column vector ${\bf p}(t) = (p(1,t),p(2,t),\ldots,p(L^2,t))^T$ ($T$ denotes transposition), and the entries of the generator $\mathbb{W}_{n n'} = - \delta_{n n'} + \Gamma_{n\leftarrow n'}$ ($\Gamma_{n\leftarrow n'} = 0$ if $n'\notin \text{nn}_n$).  The transition rates are
\begin{equation}
\Gamma_{n\leftarrow m} = e^{\frac{{\bf E}\cdot ({\bf r}_n-{\bf r}_m)}{L}} /(e^{\frac{E_x}{L}} +e^{-\frac{E_x}{L}} +e^{\frac{E_y}{L}} +e^{-\frac{E_y}{L}}),
\label{origrates}
\end{equation}
so that $\sum_n \Gamma_{n\leftarrow m}=1$, where ${\bf r}_n$ is the position vector of site $n$ taking the lattice constant as spatial unit
. The observable of interest is the partial current $J$ across a vertical slit of length $h L$ ($0<h<1$) placed in the center of the lattice, which we take as the origin of (Cartesian) coordinates. $J$ quantifies the number of particles that traverse the slit in the rightward direction minus those that do in the leftward direction per unit of time.

\noindent {\it Microscopic large-deviation methods---}  The current $J$ follows a probability distribution that adopts a large-deviation form for long times, $P(J) \sim e^{-t \varphi(J)}$,  with rate function $\varphi(J)$, which is non-negative and equal to zero only for the average current $J= \langle J \rangle$ \cite{touchette09a}. For simplicity, in the following we will consider a horizontal field ${\bf E} = (E,0)$ giving rise [cf.~\eqref{origrates}] to an average partial current $\langle J \rangle = \rho_0 h L \tanh [E/2L]$, which in the limit of large $L$ becomes $\langle J \rangle = \rho_0 Eh/2$. To investigate the fluctuations of $J$
, one would like to find the rate function $\varphi(J)$, but this is in general a difficult task, as we are dealing with events that are exponentially unlikely in time. Instead we bias the probability distribution with a parameter $s$, obtaining a new distribution
$P_s(J) = e^{-s t J} P(J)/Z_s$, where the dynamical partition function $Z_s$ acquires for long times the following large-deviation form, 
\begin{equation}
Z_s = \int dJ\,  e^{-s t J} P(J) \sim e^{t \theta(s)}.
\label{PsZs}
\end{equation}
The scaled cumulant-generating function (SCGF) $\theta(s)$ is related to the rate function by a Legendre-Fenchel transform, $\theta(s) = - \text{min}_J [s J+ \varphi(J)]$ \cite{touchette09a}.  By choosing the appropriate value of $s$ we find the fluctuation of interest, $J=\langle J \rangle_s$ (where $\langle \cdot \rangle_s$ is the average over $P_s(J)$), which in turn corresponds to (minus) the first derivative of the SCGF, $\langle J \rangle_s=-\theta'(s)$. In general, the $p$-th derivative of the SCGF yields the partial-current cumulant  of the corresponding order $\langle \langle J^p \rangle \rangle_s$:
$\lim_{t\to\infty} t^{p-1} \langle \langle J^p \rangle \rangle_s = (-1)^p \frac{d^p \theta(s)}{d s^p}$ \cite{garrahan09a}.

The SCGF can be obtained as the largest eigenvalue of the so-called tilted generator $\mathbb{W}^s$, whose entries are  \cite{touchette09a}
\begin{equation}
\mathbb{W}^s_{n n'} = \begin{cases} 
       e^{-s} \mathbb{W}_{n n'}  & n \in n_r\, \text{and}\, n' \in n_l, \\
       e^{s}\, \mathbb{W}_{n n'}&  n \in n_l\, \text{and}\, n' \in n_r,  \\
      \mathbb{W}_{n n'}  & \text{otherwise}.
   \end{cases}
   \label{tiltw}
\end{equation}
where $n_r$ contains all sites that are immediately to the right of the slit, and $n_l$ those that are immediately to the left. Notice that in Eq.~\eqref{tiltw}, $s=0$ corresponds to the original dynamics, while $s<0$ enhances current fluctuations larger than the average, and the opposite occurs for $s>0$. Since $\mathbb{W}^s$ is not symmetric, it has different left and right eigenvectors, which satisfy ${\bf l}_{s,i}^T \mathbb{W}^s = \lambda_i(s) {\bf l}_{s,i}^T$ and $\mathbb{W}^s {\bf r}_{s,i}= \lambda_i(s) {\bf r}_{s,i}$, respectively, where $i= 0,1,\ldots L^2-1$. The eigenvalues $\lambda_i(s)$ are ordered in decreasing magnitude of the real part, and $\lambda_{0}(s) = \theta(s)$. While it might appear natural that the tilted dynamics generated by \eqref{tiltw} must be a biased random walk, it turns out that such dynamics is not physical, as $\mathbb{W}^s$ does not conserve probability, $\sum_n \mathbb{W}^s_{nn'}\neq 0$. To recover the physical dynamics sustaining the fluctuation associated with a given $s$ a transformation is needed, namely the generalized Doob transform \cite{simon2009,popkov10a,jack2010,chetrite13a,chetrite15b},
\begin{equation}
\mathbb{W}^s_{\text{Doob}}=L_s \mathbb{W}^s L_s^{-1}-\theta(s) \mathbb{1}\,.
\label{wdoob}
\end{equation}
Here $L_s$ is the diagonal matrix whose entries are the components of the leading left eigenvector, ${\bf l}_{s,0}$, and $\mathbb{1}$ is the $L\times L$ identity matrix. The Doob generator $\mathbb{W}^s_{\text{Doob}}$ is a proper stochastic (probability conserving) generator, $\sum_n \mathbb{W}^s_{\text{Doob},nn'}= 0$. Its eigenvalues are $\lambda_{i}^D(s)= \lambda_i(s)-\theta(s)$, with left eigenvectors ${\bf l}_{s,i}^D = L_s^{-1} {\bf l}_{s,i}$ and right eigenvectors ${\bf r}_{s,i}^D = L_s {\bf r}_{s,i}$. The largest (zero) eigenvalue is associated with the stationary state ${\bf p}^s_{\text{stat}}={\bf r}_{s,0}^D$, $\mathbb{W}^s_{\text{Doob}} {\bf p}^s_{\text{stat}} = 0$ \footnote{The normalization of the Doob-generator eigenvectors is such that the largest absolute value of the  components of ${\bf l}^D_{s,i}$ is one, and $({\bf l}^D_{s,i})^T {\bf r}^D_{s,i} = 1$. Therefore, ${\bf l}^D_{s,0} = (1,1,\ldots,1)^T$ and the sum of all the components of ${\bf r}_{s,0}^D$ is one (${\bf p}^s_{\text{stat}} = {\bf r}_{s,0}^D$ is a probability vector).}. The Doob generator \eqref{wdoob} provides the stochastic process whose long-time statistics of $J$ is given by $P_s(J)$. 

\begin{figure*}[t!]
\includegraphics[scale=0.22]{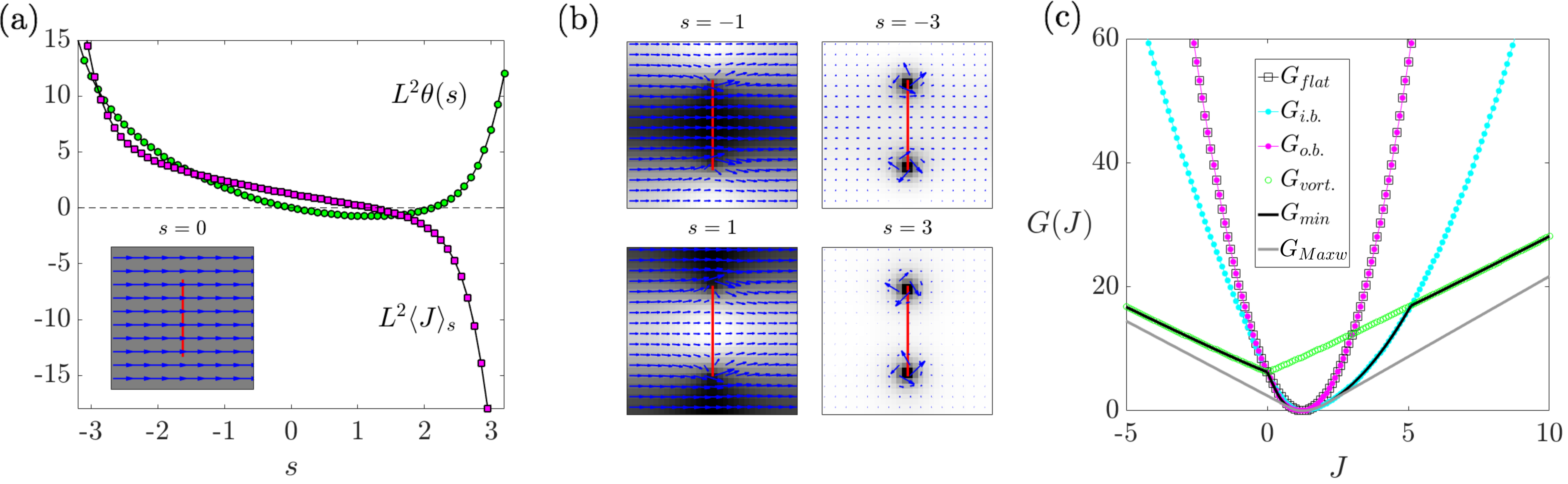}
\caption{{\sf \bf Microscopic and macroscopic large-deviation analysis of the partial current statistics in the two-dimensional random walk.} (a) ({\it Main}) Rescaled SCGF $L^2 \theta(s)$ (green circles) and average partial current $L^2 \langle J\rangle_s$ (magenta squares) for linear size $L = 32$, relative slit length $h=1/2$ and field strength $E=5$. ({\it Inset}) Density field $\rho^s(n)$ (gray color map, darker color indicates a larger number of particles) and current field ${\bf {\mathbb j}}^s(n)$ (blue arrows) for $s=0$ (natural dynamics). The slit is highlighted in red. (b) Density field $\rho^s(n)$ and current field ${\bf {\mathbb j}}^s(n)$ for $s = -1$ (corresponding to $L^2\langle J\rangle_s = 2.42$), $s = -3$ ($L^2\langle J\rangle_s = 13.09$), $s = 1$ ($L^2\langle J\rangle_s = 0.23$), and $s = 3$ ($L^2\langle J\rangle_s = -20.33$), with format as in the inset of panel (a). (c) Macroscopic fluctuation theory functional for the flat solution $G_\text{flat}(J)$, the inner-band solution  $G_\text{i.b.}(J)$, the outer-band solution $G_\text{o.b.}(J)$  and the vortex solution $ G_\text{vort.}(J)$. The minimum value among them corresponding to each $J$, denoted $G_\text{min.}(J)$, is also shown, as well as its convex envelope $G_\text{Maxw.}(J)$. See text for definitions, and the appendices for details. }
\label{Fig1}
\end{figure*}


For one random walk sustaining the fluctuation $J$, the density and current fields are thus given by ${ p}^s_{\text{stat}}(n)$ (which are the entries of ${\bf p}^s_{\text{stat}}$) and ${\bf j}^s(n)=(j^s_x(n),j^s_y(n))$. The latter satisfies $\sum_{n\in n_l} j_x^s(n)=J$, and can be obtained from the difference between the probability fluxes of two adjacent sites, $j^s_{\alpha}(n)=p^s_{\text{stat}}(n)\mathbb{W}^s_{\text{Doob},mn}-p^s_{\text{stat}}(m)\mathbb{W}^s_{\text{Doob},nm}$, where $m$ is the right (upper) neighbor of $n$ for $\alpha=x$ ($\alpha=y$). For $N$ particles, the density field, namely the average number of particles per site, is just $\rho^s(n) = N p^s_{\text{stat}}(n)$, and the current field is $ {\bf {\mathbb j}}^s(n) = N {\bf j}^s(n)$. The SCGF must also be mutiplied by $N$, as can be seen from the stochastic independence of different random walks, which yields $Z_s^N$ as the dynamical partition function, $Z_s$ being the one-particle partition function (\ref{PsZs}). 

\noindent {\it Microscopic analysis of dynamical regimes---}  In Fig.~\ref{Fig1} (a) and (b) we provide a first glimpse of the fluctuations of the partial current $J$ in a system of linear size $L = 32$ with a slit of relative length $h=1/2$ under a field of strength $E=5$. In the main panel of Fig.~\ref{Fig1}  (a) we show the (rescaled) SCGF $L^2 \theta(s)$ and the corresponding current fluctuation, $L^2 \displaystyle \langle J\rangle_s$ for different values of the biasing field $s$. 
Since $N= \rho_0 L^2$, where $\rho_0$ is the total density, we represent $\theta(s)$ and derived quantities multiplied by $L^2$, which corresponds to fixing the density to $\rho_0=1$, and will be useful in the comparison of results for different size $L$ later. When a comparison is made for a fixed number of particles $N$ instead, we shall focus on one-particle quantities ($N=1$), but then of course the density will vary as $\rho_0 = 1/L^2$.

For $s=0$ the density and the current fields are uniform, see inset of Fig.~\ref{Fig1}  (a), and the partial current is $L^2 \langle J\rangle_{s=0} = \langle J\rangle = 1.25$, which corresponds to the expression given above for $\rho_0 = 1$. For $s>0$, $L^2 \langle J\rangle_s$ decreases monotonically slowly up to $s\approx 2$, and then more rapidly, while a similar but opposite behavior is found for $s<0$. The density and current fields $\{\rho^s(n),{\bf {\mathbb j}}^s(n)\}$ for $s\neq 0$ are displayed in Fig.~\ref{Fig1} (b). 

For negative tiltings, $s<0$, 
the density becomes higher in the horizontal band crossing the slit, i.e.~$y \in (-h L /2,h L/2)$, as illustrated for $s=-1$. This regime, which will be referred to as the {\it inner band}, persists for a range of negative values of $s$, with a spatial decay for $|y|>h L/2$ that becomes more abrupt as the absolute value of $s$ increases. The band is not perfectly uniform in the horizontal direction, however, as there is a tendency for particles to populate the edges of the slit and to move cyclically around them ---an effective way to increase the partial current $J$--- which is also enhanced as the absolute value of $s$ increases. Beyond a certain point, the band practically disappears, and the (large) current is almost completely sustained by vortices localized at the edges, as shown for $s=-3$. Such cyclic, localized behavior, which has recently been observed in random walks on general graphs \cite{gutierrez2021} and the zero-range process on a diamond lattice \cite{villavicencio12a}, presents some similarities to the vortices of the two-dimensional simple exclusion process \cite{bodineau08a}. 

For positive and moderately large values of $s$ the relatively low $J$ is achieved by forming a band that avoids passing through the slit, the bulk of the density showing in the region where $|y|>h L/2$, which we will call the {\it outer band}, and is illustrated for $s=1$. But also here there is a tendency of particles to populate the edges of the slit, and move in circles (this time in the opposite direction, so the current decreases with the adopted sign convention). For sufficiently large $s$, as illustrated for $s=3$, the band again practically disappears, leading to a vortex dynamics of opposite vorticity to that observed for negative $s$. This phenomenology will be shown to be associated with a DPT to a localized state. But before addressing its nature, a macroscopic analysis will provide further insight into the dynamical regimes at play.

\noindent {\it Macroscopic analysis of dynamical regimes---}  Following the macroscopic fluctuation theory, which studies fluctuations of dynamical observables at the macroscopic (coarse-grained) level \cite{bertini15a}, we consider the probability of any trajectory of duration $T$ given by the density and current fields $\{\rho({\bf r},t),{\bf j}({\bf r},t)\}_0^{T}$. For driven diffusive systems such probability adopts a large-deviation form $P[\{\rho({\bf r},t),{\bf j}({\bf r},t)\}_0^{T}] \sim e^{- L^2 I[\rho, {\bf j}]}$, with the rate functional
\begin{equation}
 I[\rho, {\bf j}] =\!\int_0^T\!dt\!\int_\Lambda\!d{\bf r}\, \frac{\left[ {\bf j} + D(\rho) {\bf\nabla} \rho - \sigma(\rho) {\bf E} \right]^2}{2\sigma(\rho)}.
\label{MFTfunct}
\end{equation}
For a random walk, the diffusivity is $D(\rho) = 1/4$  and the mobility  $\sigma(\rho) = \rho/2$ \cite{spohn12a}. A diffusive rescaling of the microscopic variables, whereby time is rescaled by $1/L^2$ and space by $1/L$, so that the process takes place over $\Lambda = [-1/2,1/2]\!\times\![-1/2,1/2]$,  is applied. From Eq.~\eqref{MFTfunct} we can calculate by contraction the probability of any observable depending on the trajectory. Thus the probability of having a current $J$ through the slit reads $P(J)\sim\exp\{-TL^2 G(J)\}$, with $G(J)=\lim_{T\to \infty}\frac{1}{T}\min^*_{\{\rho,{\bf j}\}_0^T}I[\rho,{\bf j}]$. Here $*$ means that the minimization is subject to the constraints $J=T^{-1}\int_0^T dt \int_{-h/2}^{h/2} dy j_x(0,y;t)$ and $\partial_t \rho =-\nabla \cdot {\bf j}$ 
(continuity equation). Notice that the probability is maximized [$G(J)$ is zero] when ${\bf j}({\bf r},t)$ equals the macroscopic average current ${\bf j} = - D(\rho) {\bf\nabla} \rho + \sigma(\rho) {\bf E}$.

Since solving the two-dimensional spatiotemporal variational problem for density and current fields such as those displayed in Fig.~\ref{Fig1} (b) is a daunting task, we shall focus on a few idealized {\it ans\"atze} for $\{\rho({\bf r}), {\bf j}({\bf r})\}_0^T$. Both their time independence and their structural features are based on the previous microscopic results.
Specifically, we consider four dynamical regimes, each one leading to a different form of $G(J)$:  (i) a flat solution (uniform density and current fields) $G_\text{flat}(J)$, (ii) an inner band solution (uniform density over the region of $\Lambda$ where $|y|<h/2$) $G_\text{i.b.}(J)$, (iii) an outer band solution (uniform density over the region of $\Lambda$ where $|y|>h/2$) $G_\text{o.b.}(J)$ and (iv) a vortex solution  $G_\text{vort.}(J)$, all of which are discussed in the appendices. 
Fig.~\ref{Fig1} (c) shows $G_\text{min.}(J) = \text{min}\{G_\text{flat}(J), G_\text{i.b.}(J)$, $ G_\text{o.b.}(J),G_\text{vort.}(J)\}$, which maximizes the probability at each $J$, yielding
\begin{equation}
G_\text{min.}(J) = \begin{cases} 
          G_\text{vort.}(J) & J<0, \\
          G_\text{o.b.}(J)  & 0\leq J<\langle J\rangle,\\
          G_\text{flat}(J)  & J = \langle J\rangle,\\
          G_\text{i.b.}(J) & \langle J\rangle < J < \tilde J, \\
          G_\text{vort.}(J)   & \tilde J \leq J.\\
       \end{cases}
\end{equation}
While the flat solution maximizes the probability for $J=\langle J\rangle$, its Gaussian fluctuations around $\langle J \rangle$ are suppressed exponentially in time with respect to those of the bands:  Immediately to the right of $\langle J \rangle$ the dominant regime is the inner band, while the outer band prevails for values of $J$ between zero and $\langle J \rangle$. For larger fluctuations, $G_\text{vort.}(J)$ dominates for $J<0$, ---counterclockwise rotation in the upper edge and clockwise rotation in the lower edge of the slit--- and also for $J > \tilde{J}$ ---with vortices rotating in the opposite directions---, where $\tilde{J}$ is to an extent dependent on the choice of a cutoff radius (see appendices). 

\begin{figure}[t!]
\includegraphics[scale=0.2]{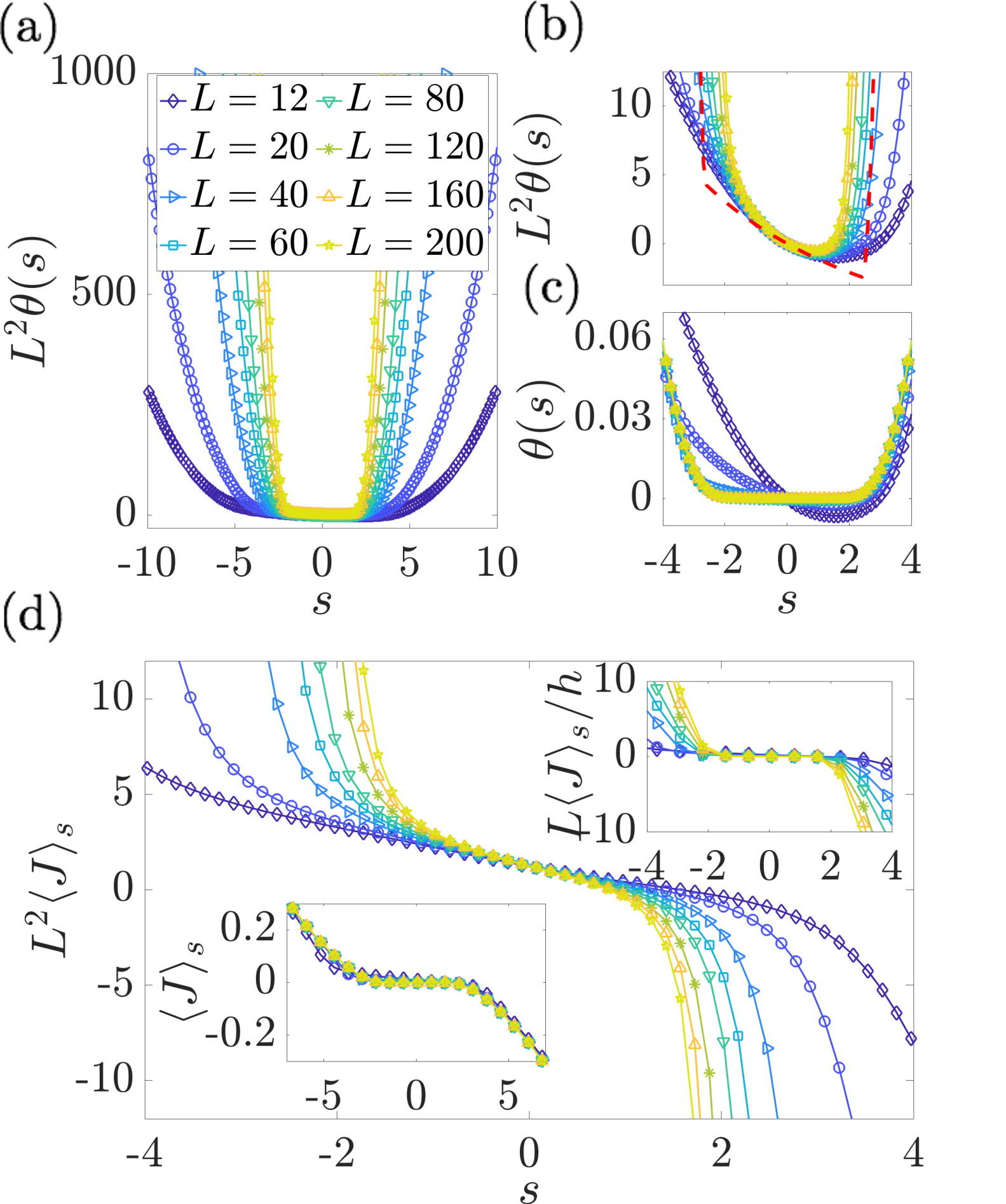}
\caption{{\sf \bf Characterization of the DPT: SCGF and average current.} (a) Density-rescaled SCGF  $L^2 \theta(s)$. (b) Same content as in panel (a) but enlarged around small values of $|s|$ and also including macroscopic SCGF (red discontinuous line). (c) SCGF $\theta(s)$. (d) Average current $\langle J\rangle_s$ multiplied by $L^2$ (main panel), by $L/h$ (top right inset) and without rescaling (lower inset). See text for justifications of rescalings. In all panels the colors and symbols represent different system sizes $L$ according to the legend in panel (a).}
\label{Fig2}
\end{figure}

While a Legendre-Fenchel transformation of $G_\text{min.}(J)$ yields the macroscopic SCGF of Fig.~\ref{Fig2} (b) (red discontinuous line; see the discussion below), an inverse Legendre-Fenchel transformation applied on such SCGF gives the convex envelope $G_\text{Maxw.}(J)$, which, unlike $G_\text{min.}(J)$ itself, is convex, see Fig.~\ref{Fig1} (c). This Maxwell construction highlights a coexistence between dynamical phases \cite{touchette09a}, namely, that of vortices and bands, specifically the outer band for $J<\langle J\rangle$ and the inner band for $J>\langle J\rangle$. Though this is broadly in agreement with the microscopic results of Fig.~\ref{Fig1} (b), it should not be taken as indicative of the existence of a first-order DPT, as $G_\text{min.}(J)$ is restricted to a few idealized cases, and is thus an upper bound of the actual macroscopic functional: Such coexistence may well approximate dynamical regimes not fully describable in terms of the {\it ans\"atze} under consideration. 

\noindent {\it Microscopic analysis of the DPT---} In order to elucidate the nature of the DPT we again focus on the microscopic SCGF $\theta(s)$. In Fig.~\ref{Fig2} (a) we show the (rescaled) SCGF $L^2 \theta(s)$ for different linear sizes ranging from $L=12$ to $200$.  A good collapse of $L^2 \theta(s)$ curves is found for $|s| \lesssim 2$, which marks the onset of the DPT, as can be better appreciated in panel (b), which also includes the macroscopic SCGF discussed in the previous section (red discontinuous line). Such an agreement, however, does not exist for $|s|$ larger than critical, as $L^2 \theta(s)$ then grows without bound as $L$ increases, yielding a rate function $\varphi(J)$ that becomes linear [qualitatively similar to $G_\text{Maxw.}(J)$ in Fig.~\ref{Fig1} (c)], and therefore an asymptotically exponential  distribution $P(J)$ \cite{touchette09a}.  Nevertheless, a good collapse is found beyond the critical points if the rescaling of $\theta(s)$ by $L^2$ is not performed (i.e.~if the number of particles $N$, instead of the density $\rho_0$, is kept fixed), as shown in panel (c). Similar conclusions can be drawn from the average current $\langle J \rangle_s$, which is shown in rescaled form $L^2 \langle J \rangle_s$ in the main panel of Fig.~\ref{Fig2} (d). After further normalization by the slit length $h L$, $L \langle J \rangle_s/h$, displayed in the upper inset, shows a good collapse below the critical points, but certainly not for larger absolute values of $s$. There the collapse is observed in the unrescaled form $\langle J \rangle_s$ (for fixed particle number $N$, without normalizing by the slit size), as shown in the lower inset.

We conclude that a collapse for different sizes $L$ occurs below the critical point for fixed density $\rho_0$, while on the other side of the transition it is the number of particles $N$ that must be kept constant. Viewed from the perspective of the idealized regimes of the previous section, the current associated with band dynamics is proportional to $\rho_0$, while that of localized vortices is proportional to the number of particles $N$.

\begin{figure}[t!]
\includegraphics[scale=0.2]{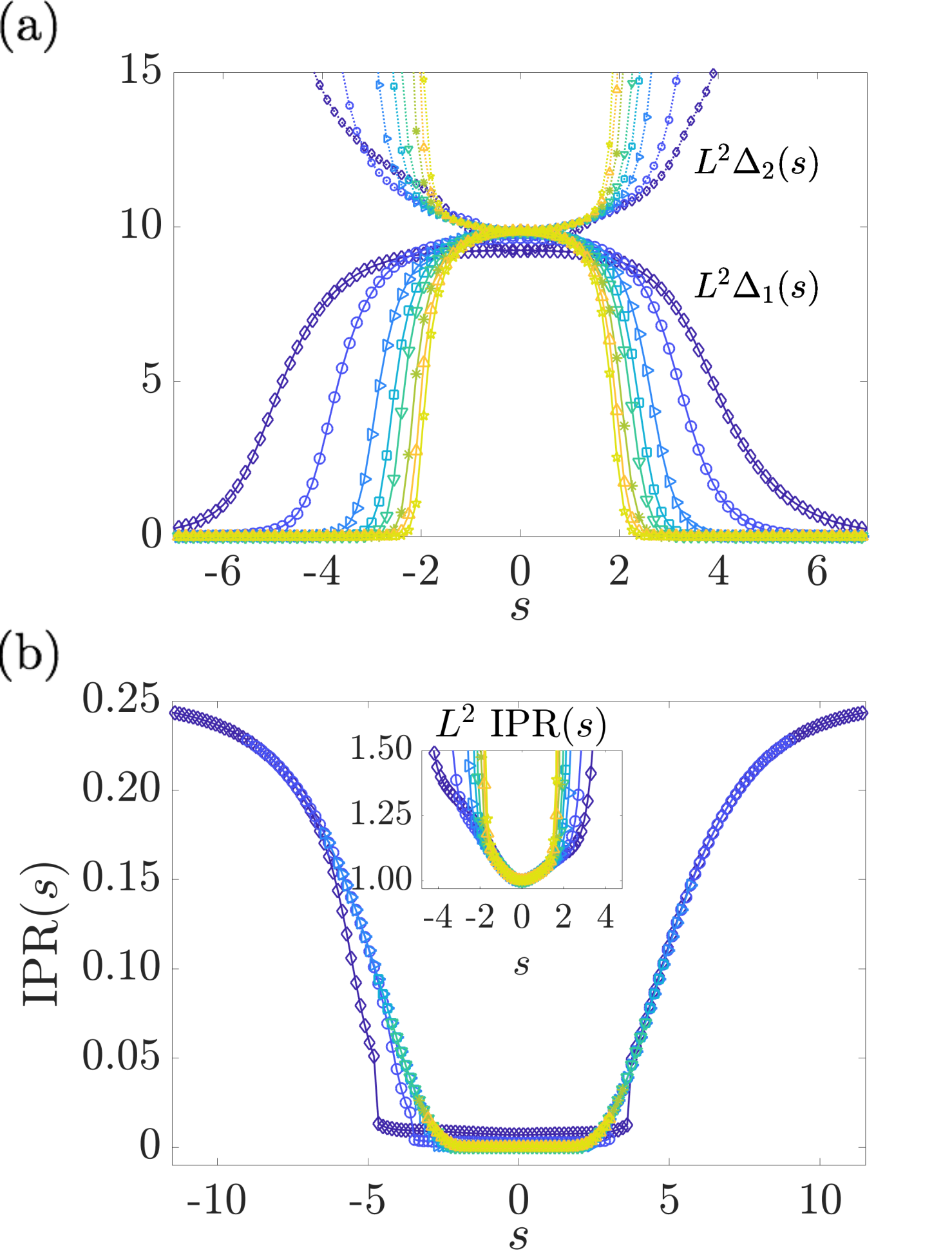}
\caption{{\sf \bf Characterization of the DPT: spectral gaps and inverse participation ratio.} (a) Rescaled spectral gaps $L^2 \Delta_1(s)$ (continuous lines, large symbols) and $L^2 \Delta_2(s)$ (dotted lines, small symbols), see text for definitions. (b) Inverse participation ratio, without (main panel) and with rescaling (inset). As $L$ increases, the range of $s$ explored decreases, due to convergence issues of the eigenvalue algorithm. In all panels the colors and symbols represent different system sizes $L$ according to the legend in Fig 2 (a).}
\label{Fig3}
\end{figure}

The features of the DPT can be inferred from the spectral properties of the Doob generator $\mathbb{W}^s_{\text{Doob}}$ \cite{gaveau06a,minganti18a,perez-espigares18a,hurtado-guti20a}, in particular from the analysis of the spectral gaps $\Delta_i(s) = \theta(s) - \text{Re}[\lambda_i(s)] = -\text{Re}[\lambda_{i}^D(s)]$ (by definition $\Delta_0(s) = 0$). The first two, $\Delta_1(s)$ (usually referred to as {\it the} spectral gap) and $\Delta_2(s)$, are shown in rescaled form (multiplied by $L^2$ so as to fix the density, $\rho_0 = 1$) in Fig.~\ref{Fig3} (a).
While $L^2 \Delta_2(s)$ only grows with $|s|$, we observe that the gap $L^2 \Delta_1(s)$ closes at a certain value, which accumulates around the critical region at the onset of the DPT. As $\lambda_1^D(s)$ is real across the range of $s$ considered (unlike $\lambda_2^D(s)$, which is complex for some values of $s$), the closing of the gap means that the stationary distribution becomes degenerate for $|s|$ above the critical value in the limit $L\to\infty$. For $|s|$ below the critical value, the stationary probability ${\bf p}^s_{\text{stat}}={\bf r}_{s,0}^D$ is unique and one can check numerically that it has a reflection symmetry $y\rightarrow -y$ also possessed by the generator: $\mathcal{U}\, \mathbb{W}^s_{\text{Doob}} \mathcal{U}^T = \mathbb{W}^s_{\text{Doob}}$, $\mathcal{U}\, {\bf p}^s_{\text{stat}}= {\bf p}^s_{\text{stat}}$, where $\mathcal{U}$ is the permutation matrix that maps the lattice site at $(x,\pm y)$ into the one at $(x,\mp y)$.
For the argument that follows, it is useful to decompose the stationary probability into its upper and lower halves, as ${\bf p}^s_{\text{stat}} = ({\bf p}^s_{\uparrow}+ {\bf p}^s_{\downarrow})/2$, where ${\bf p}^s_{\uparrow}$ (${\bf p}^s_{\downarrow}$) is zero for  $y<0$ ($y> 0$); both ${\bf p}^s_{\uparrow}$ and ${\bf p}^s_{\downarrow}$ are normalized,  and $\mathcal{U}\, {\bf p}^s_{\uparrow}= {\bf p}^s_{\downarrow}$ ($\mathcal{U}\, {\bf p}^s_{\downarrow}= {\bf p}^s_{\uparrow}$). Beyond the critical point, the right eigenspace (corresponding to the space of stationary distributions) is spanned by ${\bf r}_{s,0}^D= ({\bf p}^s_{\uparrow}+ {\bf p}^s_{\downarrow})/2$ (which by continuity shares the symmetries of ${\bf p}^s_{\text{stat}}$ below the critical value), and an orthogonal stationary state numerically found to be ${\bf r}_{s,1}^D= ({\bf p}^s_{\uparrow}- {\bf p}^s_{\downarrow})/2$, for which $\mathcal{U}\, {\bf r}_{s,1}^D= -{\bf r}_{s,1}^D$. More conveniently, one can use ${\bf p}^s_{\uparrow}$ and ${\bf p}^s_{\downarrow}$, which concentrate (more and more sharply as $|s|$ is increased) around the upper and lower edges of the slit, respectively, as a basis to form possible stationary distributions.  A particle starts from an initial site, moves towards one of the slit edges, and remains there, as the probability of making the journey to the other edge is vanishingly small for $L\to\infty$. The $\mathbb{Z}_2$ reflection symmetry is spontaneously broken.

To conclude our exploration of the DPT, we use the inverse participation ratio (IPR) as an order parameter. This is defined as the sum of the squared probabilities of occupation in the stationary state  over all sites, $\text{IPR}(s) = \sum_n [p^s_\text{stat}(n)]^2$, and is widely employed in the characterization of localized states \cite{thouless1974}. In idealized cases where the probability is evenly distributed over $\mathcal{N}$ sites, and is zero outside, the IPR is $1/\mathcal{N}$. For $s=0$ (flat solution), $\text{IPR}(0) = 1/L^2$, as shown in Fig.~\ref{Fig3} (b). In the case of infinitely sharp bands, we would have $\text{IPR} = 1/ h L^2$ for the inner band [$\text{IPR} = 1/ (1-h) L^2$ for the outer band], therefore band dynamics correspond to IPRs between $1/L^2$ and $1/h L^2$ [$1/ (1-h) L^2$], and larger values must arise from localized vortices. Accordingly, we set $\max \left\{1/ h L^2, 1/ (1-h) L^2\right\}$ as a threshold, then use ${\bf p}^s_{\text{stat}}$ to calculate the IPR for values that are smaller than such threshold, and otherwise base the calculation of the IPR on ${\bf p}^s_{\uparrow} = {\bf r}_{s,0}^D+ {\bf r}_{s,1}^D$ (or  ${\bf p}^s_{\downarrow}= {\bf r}_{s,0}^D- {\bf r}_{s,1}^D$, as they give equivalent results), as a particle is trapped in one slit edge. The argument works in the  limit of large $L$, but we use our finite-size results as approximations to that limit. The IPR shown in the main panel of Fig.~\ref{Fig3} (b) reaches a small and stable value for $s$ below the critical point. As expected from the characteristics of band dynamics, multiplication of the IPR by $L^2$ in this range of $s$ leads to an excellent collapse, see the inset of Fig.~\ref{Fig3} (b).  For large values of $|s|$, the IPR approaches a constant value of $1/4$ as $|s|\to\infty$. The latter ($\mathcal{N} = 4$) corresponds to the existence of a vortex around an edge slit, where the loop comprises four lattice sites. If instead of the stationary state ${\bf p}^s_{\uparrow}$ (or  ${\bf p}^s_{\downarrow}$) we naively take ${\bf r}_{s,0}^D$, the IPR approaches a value of $1/8$ for large $|s|$, corresponding to the existence of two vortices at the edges ($\mathcal{N} = 8$), as in the right panels of  Fig.~\ref{Fig1} (b), but sharper.

\noindent {\it Conclusions--} We have shown that the two-dimensional random walk conditioned on partial currents undergoes a DPT between delocalized bands and localized vortices, where particles condensate around the slit edges and a $\mathbb{Z}_2$ reflection symmetry is spontaneously broken. 
In principle, this DPT should be observable in experiments based on driven diffusive systems via an accurate characterization of the probability distribution of partial currents. While our analyses assume the presence of a sizable driving field, a change of field strength does not lead to any qualitatively different behavior. For zero field, however, we find that the band regime is absent from the dynamics.

Intriguing localization phenomena have also been observed in other random walks, e.g.~in the maximal entropy random walk on a lattice \cite{burda2009}, which can also be understood as the result of conditioning on a particular observable \cite{coghi2019}, or in random walks on complex networks, where some DPTs have been also investigated \cite{bacco16a}. Many interesting questions remain open, however, regarding, for example, the existence of similar localization effects when other processes (including exclusion effects or interactions) or number of spatial dimensions are considered.  Whether such effects are relevant in dissipative quantum walks \cite{attal2012,garnerone2012} (where the spectral theory of Liouvillians can be brought to bear \cite{macieszczak16a,minganti18a}) is also an interesting question, as an answer in the affirmative might stimulate a search for potential relationships to well-known localization effects in quantum mechanics \cite{anderson1958,nandkishore2015}.

\begin{acknowledgments}
We thank Rub\'en Hurtado-Guti\'errez and Pablo Hurtado for insightful discussions. The research leading to these results has received funding from the European Union's Horizon 2020 research and innovation programme under the Marie Sklodowska-Curie Cofund Programme Athenea3I Grant Agreement No.~754446, and from the Project of I+D+i Ref.~PID2020-113681GB-I00, financed by MICIN/AEI/10.13039/501100011033 and FEDER ``A way to make Europe''.  We are grateful for the the computing resources and related technical support provided by PROTEUS, the super-computing center of Institute Carlos I in Granada, Spain, and by CRESCO/ENEAGRID High Performance Computing infrastructure and its staff \cite{iannone2019}, which is funded by ENEA, the Italian National Agency for New Technologies, Energy and Sustainable Economic Development and by Italian and European research programmes.
\end{acknowledgments}

\appendix

\section{Macroscopic fluctuation theory of the two-dimensional random walk} 

The macroscopic fluctuation theory (MFT) is a powerful framework for the study of dynamical fluctuations of driven diffusive systems \cite{bertini15a}. When such processes take place over a square region of linear size $L$, the probability of a trajectory of duration $T$, $\{\rho({\bf r},t),{\bf j}({\bf r},t)\}_0^T$, adopts a large-deviation form
\begin{equation}
P[\rho({\bf r},t), {\bf j}({\bf r},t)] \sim e^{- L^2 I[\rho, {\bf j}]},
\end{equation}
where the so-called rate functional is as follows (the spatio-temporal dependence of $\rho$ and ${\bf j}$ is omitted for ease of notation)
\begin{equation}
 I[\rho, {\bf j}] =\!\int_0^T\!dt\!\int_\Lambda\!d{\bf r}\, \frac{\left[ {\bf j} + D(\rho) {\bf\nabla} \rho - \sigma(\rho) {\bf E} \right]^2}{2\sigma(\rho)}.
\end{equation}
The spatial coordinates have been rescaled so that ${\bf r} \in \Lambda = [-1/2,1/2]\!\times\![-1/2,1/2]$ (i.e.\! a square region of unit area with the origin $(0,0)$ at its center). The probability is maximized around the macroscopic average current, which in the presence of a uniform driving field {\bf E} is ${\bf j} = - D( \rho) {\bf\nabla} \rho + \sigma(\rho) {\bf E}$, where the diffusivity $D(\rho)$ and the mobility $\sigma(\rho)$ are in general functions of the density field.

\begin{figure*}
\includegraphics[scale=0.80]{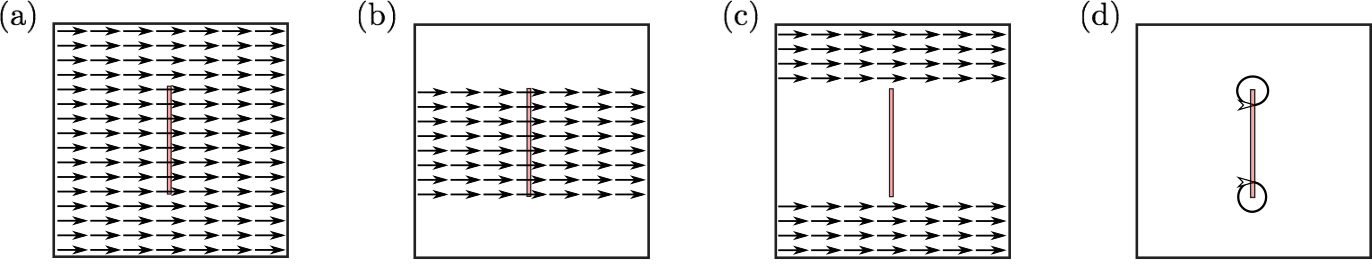}
\caption{ {\sf \bf Dynamical regimes for the two-dimensional random walk conditioned on different values of the partial current $J$ across a slit (vertical strip in the center).} (a) Flat profile. (b)  Inner band. (c) Outer band. (d) Vortices. These regimes are idealizations of those observed in the microscopic analysis based on the Doob transform.}
\label{figansatze}
\end{figure*}

For the particular case of a random walk, $D = 1/4$ and $\sigma(\rho) = \rho/2$, the functional can be expanded as
\begin{eqnarray} 
I[\rho, {\bf j}] &=&  \int_0^T\!dt \int_\Lambda\!d{\bf r}\, \frac{\left[ {\bf j} + \frac{1}{4} {\bf\nabla} \rho - \frac{1}{2}  \rho {\bf E} \right]^2}{\rho} =  \nonumber\\
&=&  \!\int_0^T\!dt\!\int_\Lambda\!d{\bf r}\, \frac{ {\bf j}^2\!+\!\frac{1}{16}  ({\bf\nabla} \rho)^2}{\rho}\!+\!\frac{ E^2}{4}\!\int_0^T\!dt\!\int_\Lambda\!d{\bf r}\,  \rho\! \\
&+&\! \frac{1}{2}   \!\int_0^T\!dt\!\int_\Lambda\!d{\bf r}\, \frac{ {\bf j}\cdot {\bf\nabla} \rho}{\rho}\!-\!\int_0^T\!dt\, {\bf E}\cdot\left(\! \int_\Lambda\!d{\bf r}\,    {\bf j}\!+\!\frac{1}{4} \!\int_\Lambda\!d{\bf r}\,   \nabla \rho\right).\nonumber
\label{preprefunctional} 
\end{eqnarray} 
As we assume the system has periodic boundary conditions, the last term is $ \int_\Lambda\!d{\bf r}\,   \nabla \rho = 0$, while the following integral can be simplified using integration by parts,
\begin{eqnarray}
\int_0^T\!dt\!\int_\Lambda\!d{\bf r}\, \frac{ {\bf j}\cdot {\bf\nabla} \rho}{\rho} &=&\int_0^T\!dt\!\int_\Lambda\!d{\bf r}\, {\bf j}\cdot {\bf\nabla} \log{\rho}\\ 
&=& \int_0^T\!dt\!\int_\Lambda\!d{\bf r}\,  {\bf\nabla}\cdot({\bf j} \log{\rho}) -  ({\bf\nabla}\cdot {\bf j})  \log{\rho}\nonumber \\
&=&   \int_0^T\!dt\!\int_\Lambda\!d{\bf r}\, (\partial_t \rho)  \log{\rho} =A(T)\!-\!A(0).\nonumber 
\label{int} 
\end{eqnarray} 
where we have taken into account that the density and current fields are coupled via the continuity equation $\partial_t \rho + \nabla \cdot {\bf j} = 0$. The function $A(t) = \int_\Lambda\!d{\bf r} (\rho \log{\rho}-\rho)$ is bounded in time, which distinguishes the corresponding term in $I[\rho, {\bf j}]$ from the other terms, as they are all time-extensive. Considering that the system is closed with a fixed number of particles $N = \rho_0 L^2$, by definition  $\rho_0 = \int_\Lambda\!d{\bf r}\,  \rho$. Together with the definition of a global current ${\bf J}_G =  \int_\Lambda\!d{\bf r}\,    {\bf j}$, all of this leads to a simpler expression for the functional, 
\begin{eqnarray} 
I[\rho, {\bf j}] &=& \int_0^T\!dt \int_\Lambda\!d{\bf r}\, \frac{ {\bf j}^2 + \frac{1}{16} ({\bf\nabla} \rho)^2}{\rho} + \frac{ E^2 T \rho_0}{4}  \nonumber \\
&+& \frac{1}{2}  (A(T) - A(0)) -   {\bf E}\cdot \int_0^T\!dt\, {\bf J}_G.
\label{prefunctional}
\end{eqnarray}
\vspace{-0.2cm}

We next calculate by contraction from the MFT functional (\ref{prefunctional}) the probability of having a current $J$ through the slit, which reads 
\begin{equation}
P(J)\sim\exp\{-TL^2 G(J)\},
\end{equation}
with $G(J)=\lim_{T\to \infty}\frac{1}{T}\min^*_{\{\rho,{\bf j}\}_0^T}I[\rho,{\bf j}]$. Here $*$ means that the minimization is subject to the constraints $J=T^{-1}\int_0^T dt \int_{-h/2}^{h/2} dy j_x(0,y;t)$ and $\partial_t \rho =-\nabla \cdot {\bf j}$. This variational problem will be applied to the different dynamical regimes discussed in the main text, which are idealizations of the  density and current fields observed in the stationary states of the Doob-transformed microscopic dynamics. Sketches of these idealized regimes are provided in Fig.~\ref{figansatze}. They all correspond to time-independent density and current fields, for which the rescaled functional can be written as
\begin{eqnarray} 
G(J)\!&=&\!\text{min}^*_{\{\rho,{\bf j}\}}\!\left[ \int_\Lambda\!d{\bf r}\, \frac{ {\bf j}^2\!+\!\frac{1}{16} ({\bf\nabla} \rho)^2}{\rho}\!+\!\frac{ E^2 \rho_0}{4}\!-\!{\bf E}\cdot {\bf J}_G\right].\nonumber \\
& &
\label{functional}
\end{eqnarray}
The third term  in (\ref{prefunctional}) has been neglected, since the discussion focuses on the long-time limit, and, as we pointed out before, $\displaystyle \lim_{T\to\infty} (A(T) - A(0))/T = 0$.

Different forms of the density and current fields conditioned on a given partial current $J$ will be evaluated in each particular regime, always focusing on situations where ${\bf E}$ is horizontal and points to the right, ${\bf E} = E \hat{\bf x}$. When a parameter characterizes an idealized regime, as embodied in a given {\it ansatz} for $\{\rho({\bf r}),{\bf j}({\bf r})\}$, the parameter choice that minimizes $G(J)$ for that value of $J$ is the one that is (overwhelmingly) more likely to be observed. When different {\it ans\"atze}, each one individually optimized by the appropriate parameter choice, are compared for a given $J$, the one that yields a minimum value of $G(J)$ is the one that would be observed in practice.

\section{Flat solution}

For a given partial current $J$ across a slit of length $h$, a flat density profile that sustains a uniform current, see Fig.~\ref{figansatze} (a), takes the following values:
\begin{equation}
\rho = \rho_0,\ \ {\bf j} = \frac{J}{h} \hat{\bf x}.
\end{equation}
The time-rescaled MFT functional (\ref{functional}) takes a simple quadratic form:
\begin{eqnarray} 
 G_\text{flat}(J) &=&  \frac{J^2}{h^2 \rho_0} + \frac{E^2 \rho_0}{4} -  \frac{E J}{h}  = \rho_0 \left(\frac{J}{h \rho_0}  - \displaystyle\frac{E}{2} \right)^2.
\end{eqnarray}
The fluctuations around the mean value $\displaystyle \langle J\rangle = \frac{E h \rho_0}{2}$ are thus Gaussian, with a variance that decreases as $1/T$,
\begin{equation}
P_\text{flat}(J) \sim \exp\left[-{\left(J - \langle J\rangle \right)^2/\displaystyle\left(\frac{h^2 \rho_0}{L^2 T}\right) }\right].
\end{equation}

\section{Inner band solution}

We next analyze the case when the density profile forms a horizontal band that decays for $|y| > h/2$, and the current is proportional to the density. For simplicity, we assume an exponential decay:
\begin{equation}
 \rho(x,y) = \begin{cases} 
          C\, e^{-(y-h/2)/\ell}  & y> h/2 \\
          C & |y| \leq h/2,\  \ \ {\bf j}(x,y) = j_0 \rho(x,y)  \hat{\bf x}. \\
          C\, e^{(y+h/2)/\ell}  & y< - h/2 
       \end{cases}
\end{equation}
In this case, there is one parameter in the solution, which is the characteristic length of the exponential decay $\ell$. In Fig.~\ref{figansatze} (b) this regime is illustrated for $\ell \ll h$; there the density profile as a function of $y$ approaches a step function.  Both density and current are invariant under translations along the horizontal axis.

In order to determine $C$, we perform the integral of the density over the square region $\Lambda$ and equate it to $\rho_0$:
\begin{eqnarray}
\rho_0 &=&  \int_{-1/2}^{1/2} dy  \rho(x,y) 
= C \left(2  \ell \left(1 - e^{-(1-h)/2\ell }\right)+ h\right)\nonumber\\
& \implies& C = \frac{\rho_0}{2  \ell \left(1 - e^{-(1-h)/2\ell}\right)+ h}. 
\end{eqnarray}
In the following it will be useful to divide the integrated density into two contributions, namely, that corresponding to particles passing through the slit $\rho_\text{in}(\ell)$, which is decreasing in $\ell$, and that corresponding to particles moving across the region where $|y|>h/2$, $\rho_\text{out}(\ell)$, which increases with $\ell$,
\begin{eqnarray}
\rho_\text{in}(\ell) &=& C h =  \frac{\rho_0 h}{2  \ell \left(1 - e^{-(1-h)/2\ell}\right)+ h},\nonumber \\ 
\rho_\text{out}(\ell) &=& \rho_0-\rho_\text{in}(\ell) = \frac{\rho_0 2  \ell \left(1 - e^{-(1-h)/2\ell}\right)}{2  \ell \left(1 - e^{-(1-h)/2\ell}\right)+ h}.\ \  \
\end{eqnarray}
As for the proportionality constant $j_0$, it can be obtained from the fact that the current integrated over the slit is by definition $J$:
\begin{eqnarray}
J =  \int_{-h/2}^{h/2} dy\,  j(x,y) = j_0 \rho_\text{in}(\ell) , \ \ \ \ j_0 = \frac{J}{\rho_\text{in}(\ell) }.\ \ \
\end{eqnarray}

The rate functional (\ref{functional}) thus takes the form
\begin{eqnarray} 
G_\text{i.b.}(J,\ell)\!&=&\!\int_\Lambda\!d{\bf r}\, \frac{j_0^2 \rho^2\!+\!\frac{1}{16} ({\partial_y} \rho)^2}{\rho} + \frac{ E^2 \rho_0}{4}-E j_0 \rho_0\nonumber\\
&=&   \rho_0 \left( \frac{J}{\rho_\text{in}(\ell) }\!-\!\frac{E}{2} \right)^2\!+\frac{\rho_\text{out}(\ell) }{16 \ell^2}.
\label{Ginn}
\end{eqnarray} 
The first term quantifies the cost of having a density $\rho_\text{in}(\ell)$ different from $2 J/E$, and the second that of forming a non-uniform density field. The flat solution previously discussed corresponds to $\rho_\text{in}(\ell) =\rho_0 h$, hence the first term is minimized for $\displaystyle J = E h \rho_0/2 =  \langle J\rangle$ in that case. As the density is then uniform, $\ell\to\infty$, the second term proportional to $\rho_\text{out}(\ell)$ also vanishes. 

Of all possible band profiles, which are characterized by different values of the characteristic length $\ell$, the one that is most likely to be observed for a given $J$ is the one that minimizes the functional. For fluctuations around the mean value, we have two possibilities: 
\begin{itemize}
\item  if $\displaystyle J >  \langle J\rangle$, a band is formed, whose vertical profile is steeper as $J$ increases, as, in order to minimize the first term, we need some $\rho_\text{in}(\ell) > \rho_0 h$, which requires having a finite value of $\ell$ (non-flat solution), so   $\rho_\text{out}(\ell)  <\rho_0 (1-h)$;
\item if $0<\displaystyle J \leq  \langle J\rangle$, we still have a flat solution, as $\rho_0 h$ is the lowest value that $\rho_\text{in}(\ell)$ can achieve, and moreover the second term vanishes for such flat solution, $\ell\to\infty$.
\end{itemize}
For $\displaystyle J >  \langle J\rangle$ there is an effective competition between the first and the second term in Eq.~(\ref{Ginn}), as the second, $\rho_\text{out}(\ell)/16 \ell^2$, includes the cost of creating a non-uniform density profile, and is a decreasing function of $\ell$. In the limit of very large $J$, the first term clearly dominates and the system adopts an abrupt vertical profile, $\ell\approx 0$, but intermediate cases between $\langle J\rangle$ and such large values of $J$ must be determined from a minimization of $G_\text{i.b.}(J,\ell)$.

\begin{figure*}
\includegraphics[scale=0.20]{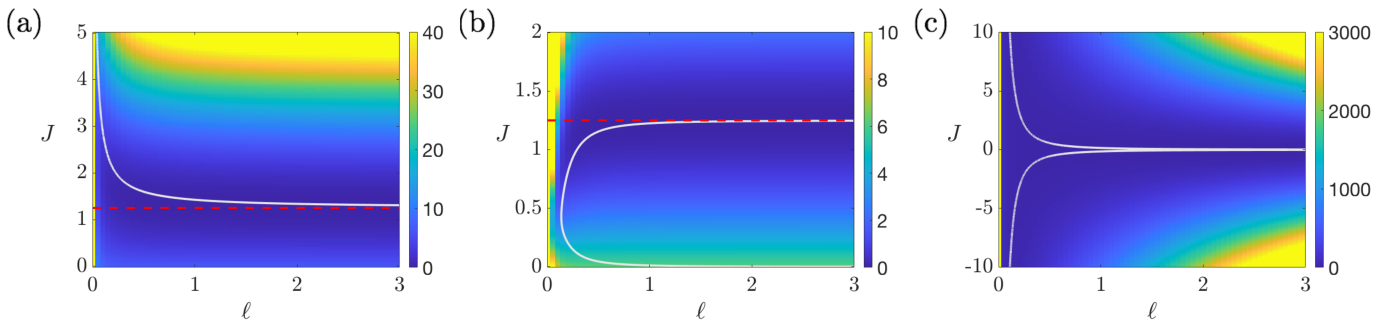}
\caption{ {\sf \bf Macroscopic functionals for the inner band, the outer band, and the vortex solution as functions of the characteristic length $\ell$ and the partial current $J$.}
(a) Inner-band functional $G_\text{i.b.}(J,\ell)$. (b) Outer-band functional $G_\text{o.b.}(J,\ell)$. (a) Vortex functional $G_\text{vort.}(J,\ell)$. In each case, the location of the minimum value of the functional for each $J$ is shown as a gray continuous line. In panels (a) and (b) the red discontinuous line corresponds to the average partial current $\langle J\rangle$.}
\label{FigS2}
\end{figure*}

Given in terms of $\ell$, Eq.~(\ref{Ginn}) is too cumbersome for a convenient analytical minimization. Instead, the values of $\ell$ that minimize this function have been found numerically. In Fig.~\ref{FigS2} (a) we provide a surface plot showing $G_\text{i.b.}(J,\ell)$ as a function of $J$ and $\ell$ for $E=5$ and $h=1/2$. The values of $\ell$ where the minimum is achieved for each $J$ are displayed as a gray continuous line. As expected, it turns out that for $0<\displaystyle J \leq  \langle J\rangle$ the minimum is achieved for $\ell \to \infty$, and for $\displaystyle J >  \langle J\rangle$ the position of the minimum gets closer and closer to zero as $J$ increases. $\langle J\rangle$ is highlighted by a red discontinuous line.

\section{Outer band solution}

We next consider the case when the density profile forms a horizontal band that avoids passing through the slit, which is constant for $|y| > h/2$ and decays for $|y| \leq h/2$,  see Fig.~\ref{figansatze} (c), where this regime is illustrated for $\ell \ll h$. As in the previous case, the current is taken to be proportional to the density. For simplicity, we again assume an exponential decay:
\begin{eqnarray}
 \rho(x,y) &=& \begin{cases} 
          C  & y> h/2 \\
         \displaystyle \frac{C}{1+e^{-h/\ell}} \left(e^{(y-h/2)/\ell} + e^{-(y+h /2)/\ell}\right) & |y| \leq h/2  \\
          C  & y< - h/2 
       \end{cases}\nonumber\\
 {\bf j}(x,y) &=& j_0 \rho(x,y)  \hat{\bf x}. 
\end{eqnarray}
where the prefactor of the middle term has been chosen so as to ensure continuity. In order to determine $C$, we need to normalize the integral of the density over the square region $\Lambda$ and equate it to $\rho_0$:
\begin{eqnarray}
\rho_0 &=&  \int_{-1/2}^{1/2} dy  \rho(x,y)  
 = C \! \left(\! 1\! -\! h\! +\!  2  \ell\!  \tanh\left(\frac{h}{2 \ell}\right)\! \right)\nonumber \\ 
&\implies & C \! =\!  \frac{\rho_0}{1\! -\! h\! +\!  2 \ell \tanh\left(\frac{h}{2 \ell}\right)}. 
\end{eqnarray}
As in the previous case, it will be useful to divide the integrated density into two contributions, that corresponding to particles passing through the slit $\rho_\text{in}(\ell)$, which is now increasing in $\ell$, and that corresponding to particles moving across the region where $|y|>h/2$, $\rho_\text{out}(\ell)$, which decreases with $\ell$,
\begin{eqnarray}
 \rho_\text{out}(\ell) &=& C (1-h) = \frac{\rho_0(1-h)}{1-h+ 2 \ell \tanh\left(\frac{h}{2 \ell}\right)},\nonumber \\ 
\rho_\text{in}(\ell) &=& \rho_0 - \rho_\text{out}(\ell)  =   \frac{\rho_0 2 \ell \tanh\left(\frac{h}{2 \ell}\right)}{1-h+ 2 \ell \tanh\left(\frac{h}{2 \ell}\right)}.
\end{eqnarray}
As for the proportionality constant $j_0$, it can be determined from the fact that the current integrated over the slit is by definition $J$:
\begin{equation}
J =  \int_{-h/2}^{h/2} dy\,  j(x,y) = j_0 \rho_\text{in}(\ell),\ \ j_0 = \frac{J}{\rho_\text{in}(\ell)}.
\end{equation} 

The rate functional is in this case
\begin{widetext} 
\begin{equation} 
G_\text{o.b.}(J,\ell)\!=\!\int_\Lambda\!d{\bf r}\, \frac{j_0^2 \rho^2 + \frac{1}{16} ({\partial_y} \rho)^2}{\rho} + \frac{E^2 \rho_0}{4} -   E j_0 \rho_0 
 = \rho_0 \left(\frac{J}{\rho_\text{in}(\ell)}\!-\!\frac{E}{2}\right)^2 + \frac{ \rho_0 \left(\sinh\left(\frac{h}{2 \ell}\right)\!+\!\arctan{e^{-h/2\ell}}\!-\!\arctan{e^{h/2\ell}}\right)  }{4 \ell e^{h/2\ell}  (1+e^{-h/\ell}) (1-h+ 2 \ell \tanh\left(\frac{h}{2 \ell}\right))}.
\label{Gout}
\end{equation}
\end{widetext} 
Again, the first term quantifies the cost of having a density $\rho_\text{in}(\ell)$ different from $2 J/E$, and the second that of forming a non-uniform density field. The flat solution is given by $\rho_\text{in}(\ell) =\rho_0 h$, hence the first term is minimized for $\displaystyle J = \langle J\rangle$ in that case. This corresponds to a uniform density, $\ell\to\infty$, for which the second term also vanishes. 

Of all possible band profiles, which are characterized by different values of the characteristic length $\ell$, the one that is most likely to be observed for a given $J$, is the one that minimizes the functional. For fluctuations around the mean value, we have two possibilities: 
\begin{itemize}
\item  if $\displaystyle 0< J <  \langle J\rangle$, a band is formed, whose vertical profile is steeper as $J$ decreases from $\langle J\rangle$, at least up to certain value of $J$ (more about this later), as, in order to minimize the first term, we need some $\rho_\text{in}(\ell) < \rho_0 h$, which requires having a finite value of $\ell$ (non-flat solution), so   $\rho_\text{out}(\ell)  >\rho_0 (1-h)$;
\item if $J \geq  \langle J\rangle$, we have a flat solution, as $\rho_0 h$ is the highest value that $\rho_\text{in}(\ell)$ can achieve, and moreover the second term vanishes for such flat solution, $\ell\to\infty$.
\end{itemize}
For $\displaystyle 0< J <  \langle J\rangle$ there is an effective competition between the first and the second terms in Eq.~(\ref{Gout}), as the second includes the cost creating a non-uniform density profile, and is a decreasing function of $\ell$. As $J$ decreases from the average value $\langle J\rangle$, the first term tries to adapt to a smaller $\rho_\text{in}(\ell)$, so $\ell$ becomes smaller, at the cost of creating a non-uniform profile, which contributes to the second term of $G_\text{o.b.}(J,\ell)$. However, when $J$ gets very small, deviations around the average in the first term become less important than the second term, and therefore we again obtain a flat solution, $\ell \to \infty$.

The values of $\ell$ that minimize $G_\text{o.b.}(J,\ell)$ have been found numerically. In Fig.~\ref{FigS2} (b) we provide a surface plot showing $G_\text{o.b.}(J,\ell)$ as a function of $J$ and $\ell$ for $E=5$ and $h=1/2$. The values of $\ell$ where the minimum value is achieved for each $J$ are displayed as a gray continuous line. As expected, it turns out that for $\displaystyle J >  \langle J\rangle$ the minimum is always achieved for $\ell \to \infty$, and for $0<\displaystyle J \leq  \langle J\rangle$  the value of $\ell$ where the minimum is achieved has the non-monotonic behavior discussed in the previous paragraph. $\langle J\rangle$ is highlighted by a red discontinuous line.

\section{Vortices}

Finally, we consider a solution where vortices form around the edges of the slit,  see Fig.~\ref{figansatze} (d), which as mentioned in the main text resembles a configuration previously described in the two-dimensional simple exclusion process conditioned on partial currents \cite{bodineau08a}. Due to the rotational symmetry of the solution, the density is best represented in polar coordinates, by some decreasing function of the radial distance $r$ from the center of the vortex. We again assume the density profile to be of exponential form due to its simplicity
\begin{equation}
 \rho(r,\varphi) = \begin{cases} 
          0  & r< \tilde{r}, \\
          C e^{-(r-\tilde{r})/\ell}   & r\geq\tilde{r}. 
       \end{cases}
\end{equation}
Again the steepness of the profile is given by the characteristic length $\ell$. The role played by the cutoff radius $\tilde{r}$ will be discussed later. The normalization in this case is given by 
\begin{eqnarray} 
\rho_0 &=& 2 \int_0^{2\pi}\! d\varphi  \int_{\tilde{r}}^\infty\!dr\,  r\, C\, e^{-(r-\tilde{r})/\ell}
= 4 \pi C  \ell \left(\tilde{r} +\ell \right)\nonumber\\ 
&\implies& C =\frac{\rho_0}{4 \pi \ell \left(\tilde{r} + \ell \right)}. 
\end{eqnarray} 
The factor of $2$ multiplying the integral of the density arises from the existence of two vortices, each one located in the vicinity of each of the two edges of the slit.  The upper limit of integration in the integral over $r$ should not exceed $h/2$, but at least for $\ell \ll h$ the simplification arising from taking it to infinity is expected to compensate for any slight loss of accuracy, as we will explain below.

The current at the vortices in the microscopic results was numerically found to be roughly proportional to the density divided by $r$. For simplicity we assume a strict proportionality $j(r) \propto \rho(r)/r$ in our ansatz:
\begin{equation}
{\bf j}(r) = j_0 \frac{\rho(r)}{r} \hat{\bf \varphi}  = \begin{cases} 
          {\bf 0}  & r< \tilde{r}, \\
          \displaystyle \frac{j_0\, C\, e^{-(r-\tilde{r})/\ell}}{r}\,  \hat{\bf \varphi}  & r\geq \tilde{r}.
       \end{cases}
\end{equation}
The proportionality factor $j_0$ is as usual determined from a calculation of the partial current through the slit
\begin{eqnarray}
J &=& 2 \int_{\tilde{r}}^\infty dr j(r) 
=  2 j_0 C e^{\tilde{r}/\ell}  \Gamma(0,\tilde{r}/\ell)\nonumber \\
 &\implies& j_0  
= \frac{J 2 \pi \ell \left(\tilde{r} +\ell \right) e^{-\tilde{r}/\ell}}{\rho_0   \Gamma(0,\tilde{r}/\ell)},
\end{eqnarray}
where the final expression contains an incomplete gamma function evaluation.

The rate functional is in this case
\begin{eqnarray} 
G_\text{vort.}(J,\ell)\!&=& \int_\Lambda\!d{\bf r}\, \frac{ {\bf j}^2 + \frac{1}{16} (\partial_r \rho)^2}{\rho} + \frac{E^2\rho_0}{4} -  {\bf E}\cdot {\bf J}_G\nonumber\\
&=& 
 \frac{J^2 4 \pi^2 \ell \left(\tilde{r} +\ell \right) e^{-\tilde{r}/\ell}}{\rho_0   \Gamma(0,\tilde{r}/\ell)}\!+\!\frac{\rho_0}{8 \ell^2}\!+\!\frac{E^2\rho_0}{4}, 
\end{eqnarray}
where we have taken into consideration that, due to the vortex geometry, the total current vanishes ${\bf J}_G = {\bf 0}$.

\begin{figure*}
\includegraphics[scale=0.25]{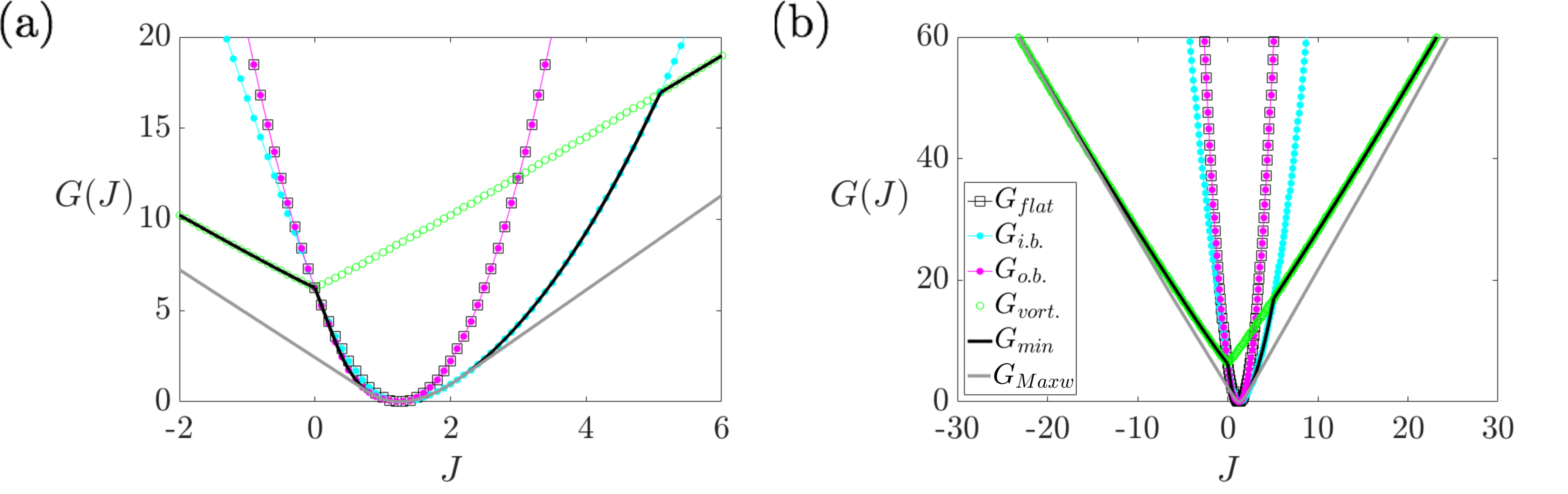}
\caption{ {\sf \bf Comparison of the MFT functional for the different dynamical regimes under consideration.}
The MFT functional of the flat solution $G_\text{flat}(J)$, the inner-band solution  $G_\text{i.b.}(J,\ell^*(J))$, the outer-band solution $G_\text{o.b.}(J,\ell^*(J))$  and the vortex solution $ G_\text{vort.}(J,\ell^*(J))$ are displayed. The minimum value among them corresponding to each $J$, which is denoted $G_\text{min.}(J)$, is also shown, as well as its convex envelope $G_\text{Maxw.}(J)$. Panels (a) and (b) contain the same results with significantly different axis ranges, thus making it possible to inspect different aspects of them (see text for an explanation). Fig.~\ref{Fig1} (c) displays the same results with yet another choice of axis ranges.}
\label{FigS3}
\end{figure*}

The values of $\ell$ that minimize $G_\text{vort.}(J,\ell)$ have been found numerically. In Fig.~\ref{FigS2} (c) we provide a surface plot showing $G_\text{vort.}(J,\ell)$ as a function of $J$ and $\ell$ for $E=5$ and $h=1/2$. The values of $\ell$ where the minimum value is achieved for each $J$ are displayed as a gray continuous line. We see that the $\ell$ that minimizes $G_\text{vort.}(J,\ell)$  decreases as $|J|$ increases. For $|J| \gtrapprox 2$, the condition $\ell \ll h$ that was assumed before in order to simplify certain integrals is indeed satisfied. In the figure we have set the cutoff $\tilde r = 0.001$, as the qualitative dependence on $\ell$ is independent of this choice. As for the quantitative influence of the cutoff $\tilde r$ on the minimum value that is achieved by $G_\text{vort.}(J,\ell)$ for a given $J$, it will be discussed next when the MFT functionals of the four dynamical regimes that have been considered (that is, flat solution, inner band, outer band, and vortices) are compared over a range of fluctuations of the partial current $J$.

\section{Comparison of different dynamical regimes}

In order to determine the most likely form of a fluctuation away from the mean, $J\neq \langle J\rangle$, the functionals corresponding to the different dynamical regimes must be compared. This includes the flat solution $G_\text{flat}(J)$, the inner-band solution  $G_\text{i.b.}(J,\ell^*(J))$, the outer-band solution $G_\text{o.b.}(J,\ell^*(J))$  and the vortex solution $ G_\text{vort.}(J,\ell^*(J))$, where $\ell^*(J)$ denotes the value of $\ell$ for which each of those functionals are minimized for a given $J$, which was highlighted by gray continuous lines in the different panels of Fig.~\ref{FigS2}. The dependence on $\ell^*(J)$ is omitted in the main text, as the role played by the characteristic length $\ell$ is not discussed there.

A comparison of the different dynamical regimes is provided in Fig.~\ref{FigS3}. Aside from the functionals corresponding to the different {\it ans\"atze}, we include the functional that takes the minimum value at each $J$, $G_\text{min.}(J) = \text{min}\{G_\text{flat}(J), G_\text{i.b.}(J,\ell^*(J)), G_\text{o.b.}(J,\ell^*(J)),$ $G_\text{vort.}(J,\ell^*(J))\}$,
and also its convex envelope $G_\text{Maxw.}(J)$. While $G_\text{min.}(J)$, from which the macroscopic SCGF of Fig.~\ref{Fig2} (b) has been obtained via a Legendre-Fenchel transform, maximizes the probability for each value of $J$,  it is not convex.  If the inverse Legendre-Fenchel transform is performed on the SCGF, we then obtain $G_\text{Maxw.}(J)$, which is convex. This Maxwell construction highlights a coexistence between the different dynamical phases \cite{touchette09a}.

The two panels of  Fig.~\ref{FigS3} show essentially the same results, which were moreover displayed in Fig.~\ref{Fig1} (c), but the axis ranges are very different, and thus focus on different aspects. In Fig.~\ref{FigS3} (a), which is meant to facilitate the comparison between the different dynamical regimes, we can clearly see that  $G_\text{min.}(J)$ takes values across the range of $J$ as follows:
\begin{equation}
G_\text{min.}(J) = \begin{cases} 
          G_\text{vort.}(J,\ell^*(J)) & J<0, \\
          G_\text{o.b.}(J,\ell^*(J))  & 0\leq J<\langle J\rangle,\\
          G_\text{flat}(J)  & J = \langle J\rangle,\\
          G_\text{i.b.}(J,\ell^*(J)) & \langle J\rangle < J < \tilde J, \\
          G_\text{vort.}(J,\ell^*(J))   & \tilde J \leq J.\\
       \end{cases}
\end{equation}
The average $\langle J\rangle$ corresponds to the absolute minimum of $G_\text{min.}(J)$ where the flat solution and the band solutions are equal, $G_\text{min.}(\langle J\rangle) = G_\text{flat}(\langle J\rangle) = G_\text{i.b.}(\langle J\rangle,\ell^*(\langle J\rangle))  = G_\text{o.b.}(\langle J\rangle,\ell^*(\langle J\rangle)),$ as expected from the discussions above. All solutions take the value $\displaystyle G_\text{flat}(0) = \rho_0 E^2/4$ for $J=0$, including $G_\text{vort.}(J,\ell^*(J))$, which dominates immediately to the left, i.e. for $J<0$, ---corresponding to clockwise rotation in the upper edge and counterclockwise rotation in the lower edge of the slit--- and also for $J \geq \tilde{J}$ ---corresponding to counterclockwise rotation in the upper edge and clockwise rotation in the lower edge of the slit---, where $\tilde{J}$ is to some extent dependent on the choice of the cutoff $\tilde r$. Indeed $G_\text{vort.}(J,\ell^*(J))$ for different values of $\tilde r$ is qualitatively similar, starting from the same value at $J=0$, and growing symmetrically to right and left. However, the steepness of the increase as $|J|$ grows decreases as $\tilde r$ is made smaller, which in $G_\text{min.}(J)$  is only reflected in a slower growth for $J<0$ or $J>\tilde{J}$, and in a leftwards shift of $\tilde{J}$.

In Fig.~\ref{FigS3} (b)  $G_\text{Maxw.}(J)$ is shown to arise from a coexistence of the outer-band solution and the vortex solution for $J<\langle J\rangle$, and from a coexistence of the inner-band solution and the vortex solution for $J>\langle J\rangle$. Such coexistences, which seem to be consistent with the superpositions of bands and vortices observed in the Doob-transformed microscopic dynamics of Fig.~\ref{Fig1} (b), should not be interpreted literally as signaling the existence of a first-order DPT in the random walks. We are considering a limited subset of idealized dynamical behaviors, and what is here described by a coexistence of them may in fact correspond to dynamical behaviors that are not fully describable in terms of the {\it ans\"atze} under consideration. In fact, our $G_\text{min.}(J)$ is an upper bound of the actual macroscopic functional restricted to a quite limited subset of density and current fields $\{\rho({\bf r}),{\bf j}({\bf r})\}$ (though qualitatively representative of some of the main features that are observed in the microscopic analysis), as is frequently the case is the analysis of mean-field solutions of dynamical large deviations of many-body systems \cite{perez-espigares18b}. Yet, despite its simplicity, our study does provide a sound qualitative understanding of the dynamical regimes under consideration ---if not of the nature of the phase transition, which is actually continuous, as explained in the main text.


\begin{thebibliography}{65}%
\makeatletter
\providecommand \@ifxundefined [1]{%
 \@ifx{#1\undefined}
}%
\providecommand \@ifnum [1]{%
 \ifnum #1\expandafter \@firstoftwo
 \else \expandafter \@secondoftwo
 \fi
}%
\providecommand \@ifx [1]{%
 \ifx #1\expandafter \@firstoftwo
 \else \expandafter \@secondoftwo
 \fi
}%
\providecommand \natexlab [1]{#1}%
\providecommand \enquote  [1]{``#1''}%
\providecommand \bibnamefont  [1]{#1}%
\providecommand \bibfnamefont [1]{#1}%
\providecommand \citenamefont [1]{#1}%
\providecommand \href@noop [0]{\@secondoftwo}%
\providecommand \href [0]{\begingroup \@sanitize@url \@href}%
\providecommand \@href[1]{\@@startlink{#1}\@@href}%
\providecommand \@@href[1]{\endgroup#1\@@endlink}%
\providecommand \@sanitize@url [0]{\catcode `\\12\catcode `\$12\catcode
  `\&12\catcode `\#12\catcode `\^12\catcode `\_12\catcode `\%12\relax}%
\providecommand \@@startlink[1]{}%
\providecommand \@@endlink[0]{}%
\providecommand \url  [0]{\begingroup\@sanitize@url \@url }%
\providecommand \@url [1]{\endgroup\@href {#1}{\urlprefix }}%
\providecommand \urlprefix  [0]{URL }%
\providecommand \Eprint [0]{\href }%
\providecommand \doibase [0]{http://dx.doi.org/}%
\providecommand \selectlanguage [0]{\@gobble}%
\providecommand \bibinfo  [0]{\@secondoftwo}%
\providecommand \bibfield  [0]{\@secondoftwo}%
\providecommand \translation [1]{[#1]}%
\providecommand \BibitemOpen [0]{}%
\providecommand \bibitemStop [0]{}%
\providecommand \bibitemNoStop [0]{.\EOS\space}%
\providecommand \EOS [0]{\spacefactor3000\relax}%
\providecommand \BibitemShut  [1]{\csname bibitem#1\endcsname}%
\let\auto@bib@innerbib\@empty
\bibitem [{\citenamefont {Bodineau}\ and\ \citenamefont
  {Derrida}(2004)}]{bodineau04a}%
  \BibitemOpen
  \bibfield  {author} {\bibinfo {author} {\bibfnamefont {T.}~\bibnamefont
  {Bodineau}}\ and\ \bibinfo {author} {\bibfnamefont {B.}~\bibnamefont
  {Derrida}},\ }\bibfield  {title} {\enquote {\bibinfo {title} {Current
  fluctuations in nonequilibrium diffusive systems: {An} additivity
  principle},}\ }\href
  {http://journals.aps.org/prl/abstract/10.1103/PhysRevLett.92.180601}
  {\bibfield  {journal} {\bibinfo  {journal} {Phys. Rev. Lett.}\ }\textbf
  {\bibinfo {volume} {92}},\ \bibinfo {pages} {180601} (\bibinfo {year}
  {2004})}\BibitemShut {NoStop}%
\bibitem [{\citenamefont {Bodineau}\ and\ \citenamefont
  {Derrida}(2006)}]{bodineau06a}%
  \BibitemOpen
  \bibfield  {author} {\bibinfo {author} {\bibfnamefont {T.}~\bibnamefont
  {Bodineau}}\ and\ \bibinfo {author} {\bibfnamefont {B.}~\bibnamefont
  {Derrida}},\ }\bibfield  {title} {\enquote {\bibinfo {title} {Current large
  deviations for asymmetric exclusion processes with open boundaries},}\ }\href
  {http://link.springer.com/article/10.1007/s10955-006-9048-4} {\bibfield
  {journal} {\bibinfo  {journal} {J. Stat. Phys.}\ }\textbf {\bibinfo {volume}
  {123}},\ \bibinfo {pages} {277--300} (\bibinfo {year} {2006})}\BibitemShut
  {NoStop}%
\bibitem [{\citenamefont {Derrida}()}]{derrida07a}%
  \BibitemOpen
  \bibfield  {author} {\bibinfo {author} {\bibfnamefont {B.}~\bibnamefont
  {Derrida}},\ }\bibfield  {title} {\enquote {\bibinfo {title} {Non-equilibrium
  steady states: fluctuations and large deviations of the density and of the
  current},}\ }\href
  {http://iopscience.iop.org/article/10.1088/1742-5468/2007/07/P07023}
  {\bibinfo  {journal} {J. Stat. Mech. P07023 (2007)}\ }\BibitemShut {NoStop}%
\bibitem [{\citenamefont {Evans}\ \emph {et~al.}(1993)\citenamefont {Evans},
  \citenamefont {Cohen},\ and\ \citenamefont {Morriss}}]{evans93a}%
  \BibitemOpen
\bibfield  {journal} {  }\bibfield  {author} {\bibinfo {author} {\bibfnamefont
  {D.~J.}\ \bibnamefont {Evans}}, \bibinfo {author} {\bibfnamefont {E.~G.~D.}\
  \bibnamefont {Cohen}}, \ and\ \bibinfo {author} {\bibfnamefont {G.~P.}\
  \bibnamefont {Morriss}},\ }\bibfield  {title} {\enquote {\bibinfo {title}
  {Probability of 2nd law violations in shearing steady-states},}\ }\href
  {http://journals.aps.org/prl/abstract/10.1103/PhysRevLett.71.2401} {\bibfield
   {journal} {\bibinfo  {journal} {Phys. Rev. Lett.}\ }\textbf {\bibinfo
  {volume} {71}},\ \bibinfo {pages} {2401--2404} (\bibinfo {year}
  {1993})}\BibitemShut {NoStop}%
\bibitem [{\citenamefont {Gallavotti}\ and\ \citenamefont
  {Cohen}(1995)}]{gallavotti95a}%
  \BibitemOpen
  \bibfield  {author} {\bibinfo {author} {\bibfnamefont {G.}~\bibnamefont
  {Gallavotti}}\ and\ \bibinfo {author} {\bibfnamefont {E.~G.~D.}\ \bibnamefont
  {Cohen}},\ }\bibfield  {title} {\enquote {\bibinfo {title} {Dynamical
  ensembles in nonequilibrium statistical mechanics},}\ }\href
  {http://journals.aps.org/prl/abstract/10.1103/PhysRevLett.74.2694} {\bibfield
   {journal} {\bibinfo  {journal} {Phys. Rev. Lett.}\ }\textbf {\bibinfo
  {volume} {74}},\ \bibinfo {pages} {2694} (\bibinfo {year}
  {1995})}\BibitemShut {NoStop}%
\bibitem [{\citenamefont {Kurchan}(1998)}]{kurchan98a}%
  \BibitemOpen
  \bibfield  {author} {\bibinfo {author} {\bibfnamefont {J.}~\bibnamefont
  {Kurchan}},\ }\bibfield  {title} {\enquote {\bibinfo {title} {Fluctuation
  theorem for stochastic dynamics},}\ }\href
  {http://iopscience.iop.org/article/10.1088/0305-4470/31/16/003/meta}
  {\bibfield  {journal} {\bibinfo  {journal} {J. Phys. A}\ }\textbf {\bibinfo
  {volume} {31}},\ \bibinfo {pages} {3719--3729} (\bibinfo {year}
  {1998})}\BibitemShut {NoStop}%
\bibitem [{\citenamefont {Lebowitz}\ and\ \citenamefont
  {Spohn}(1999)}]{lebowitz99a}%
  \BibitemOpen
  \bibfield  {author} {\bibinfo {author} {\bibfnamefont {J.~L.}\ \bibnamefont
  {Lebowitz}}\ and\ \bibinfo {author} {\bibfnamefont {H.}~\bibnamefont
  {Spohn}},\ }\bibfield  {title} {\enquote {\bibinfo {title} {A
  {Gallavotti-Cohen-type} symmetry in the large deviation functional for
  stochastic dynamics},}\ }\href
  {http://link.springer.com/article/10.1023%2FA%3A1004589714161} {\bibfield
  {journal} {\bibinfo  {journal} {J. Stat. Phys.}\ }\textbf {\bibinfo {volume}
  {95}},\ \bibinfo {pages} {333--365} (\bibinfo {year} {1999})}\BibitemShut
  {NoStop}%
\bibitem [{\citenamefont {Maes}(1999)}]{maes99a}%
  \BibitemOpen
  \bibfield  {author} {\bibinfo {author} {\bibfnamefont {C.}~\bibnamefont
  {Maes}},\ }\bibfield  {title} {\enquote {\bibinfo {title} {The fluctuation
  theorem as a {G}ibbs property},}\ }\href {\doibase 10.1023/A:1004541830999}
  {\bibfield  {journal} {\bibinfo  {journal} {Journal of Statistical Physics}\
  }\textbf {\bibinfo {volume} {95}},\ \bibinfo {pages} {367--392} (\bibinfo
  {year} {1999})}\BibitemShut {NoStop}%
\bibitem [{\citenamefont {Harris}\ and\ \citenamefont
  {Schuetz}(2007)}]{harris07a}%
  \BibitemOpen
  \bibfield  {author} {\bibinfo {author} {\bibfnamefont {R.~J.}\ \bibnamefont
  {Harris}}\ and\ \bibinfo {author} {\bibfnamefont {G.~M.}\ \bibnamefont
  {Schuetz}},\ }\bibfield  {title} {\enquote {\bibinfo {title} {Fluctuation
  theorems for stochastic dynamics},}\ }\href
  {http://iopscience.iop.org/article/10.1088/1742-5468/2007/07/P07020/meta}
  {\bibfield  {journal} {\bibinfo  {journal} {J. Stat. Mech.}\ ,\ \bibinfo
  {pages} {P07020}} (\bibinfo {year} {2007})}\BibitemShut {NoStop}%
\bibitem [{\citenamefont {Garrahan}\ \emph {et~al.}(2007)\citenamefont
  {Garrahan}, \citenamefont {Jack}, \citenamefont {Lecomte}, \citenamefont
  {Pitard}, \citenamefont {van Duijvendijk},\ and\ \citenamefont {van
  Wijland}}]{garrahan07a}%
  \BibitemOpen
  \bibfield  {author} {\bibinfo {author} {\bibfnamefont {J.~P.}\ \bibnamefont
  {Garrahan}}, \bibinfo {author} {\bibfnamefont {R.~L.}\ \bibnamefont {Jack}},
  \bibinfo {author} {\bibfnamefont {V.}~\bibnamefont {Lecomte}}, \bibinfo
  {author} {\bibfnamefont {E.}~\bibnamefont {Pitard}}, \bibinfo {author}
  {\bibfnamefont {K.}~\bibnamefont {van Duijvendijk}}, \ and\ \bibinfo {author}
  {\bibfnamefont {F.}~\bibnamefont {van Wijland}},\ }\bibfield  {title}
  {\enquote {\bibinfo {title} {Dynamical first-order phase transition in
  kinetically constrained models of glasses},}\ }\href
  {http://journals.aps.org/prl/abstract/10.1103/PhysRevLett.98.195702}
  {\bibfield  {journal} {\bibinfo  {journal} {Phys. Rev. Lett.}\ }\textbf
  {\bibinfo {volume} {98}},\ \bibinfo {pages} {195702} (\bibinfo {year}
  {2007})}\BibitemShut {NoStop}%
\bibitem [{\citenamefont {Garrahan}\ \emph {et~al.}(2009)\citenamefont
  {Garrahan}, \citenamefont {Jack}, \citenamefont {Lecomte}, \citenamefont
  {Pitard}, \citenamefont {van Duijvendijk},\ and\ \citenamefont {van
  Wijland}}]{garrahan09a}%
  \BibitemOpen
  \bibfield  {author} {\bibinfo {author} {\bibfnamefont {J.~P.}\ \bibnamefont
  {Garrahan}}, \bibinfo {author} {\bibfnamefont {R.~L.}\ \bibnamefont {Jack}},
  \bibinfo {author} {\bibfnamefont {V.}~\bibnamefont {Lecomte}}, \bibinfo
  {author} {\bibfnamefont {E.}~\bibnamefont {Pitard}}, \bibinfo {author}
  {\bibfnamefont {K.}~\bibnamefont {van Duijvendijk}}, \ and\ \bibinfo {author}
  {\bibfnamefont {F.}~\bibnamefont {van Wijland}},\ }\bibfield  {title}
  {\enquote {\bibinfo {title} {First-order dynamical phase transition in models
  of glasses: an approach based on ensembles of histories},}\ }\href
  {http://iopscience.iop.org/article/10.1088/1751-8113/42/7/075007} {\bibfield
  {journal} {\bibinfo  {journal} {J. Phys. A}\ }\textbf {\bibinfo {volume}
  {42}},\ \bibinfo {pages} {075007} (\bibinfo {year} {2009})}\BibitemShut
  {NoStop}%
\bibitem [{\citenamefont {Garrahan}\ and\ \citenamefont
  {Lesanovsky}(2010)}]{garrahan10a}%
  \BibitemOpen
  \bibfield  {author} {\bibinfo {author} {\bibfnamefont {J.~P.}\ \bibnamefont
  {Garrahan}}\ and\ \bibinfo {author} {\bibfnamefont {I.}~\bibnamefont
  {Lesanovsky}},\ }\bibfield  {title} {\enquote {\bibinfo {title}
  {Thermodynamics of quantum jump trajectories},}\ }\href
  {http://journals.aps.org/prl/abstract/10.1103/PhysRevLett.104.160601}
  {\bibfield  {journal} {\bibinfo  {journal} {Phys. Rev. Lett.}\ }\textbf
  {\bibinfo {volume} {104}},\ \bibinfo {pages} {160601} (\bibinfo {year}
  {2010})}\BibitemShut {NoStop}%
\bibitem [{\citenamefont {Carollo}\ \emph {et~al.}(2017)\citenamefont
  {Carollo}, \citenamefont {Garrahan}, \citenamefont {Lesanovsky},\ and\
  \citenamefont {P\'erez-Espigares}}]{carollo17a}%
  \BibitemOpen
  \bibfield  {author} {\bibinfo {author} {\bibfnamefont {F.}~\bibnamefont
  {Carollo}}, \bibinfo {author} {\bibfnamefont {J.~P.}\ \bibnamefont
  {Garrahan}}, \bibinfo {author} {\bibfnamefont {I.}~\bibnamefont
  {Lesanovsky}}, \ and\ \bibinfo {author} {\bibfnamefont {C.}~\bibnamefont
  {P\'erez-Espigares}},\ }\bibfield  {title} {\enquote {\bibinfo {title}
  {Fluctuating hydrodynamics, current fluctuations, and hyperuniformity in
  boundary-driven open quantum chains},}\ }\href {\doibase
  10.1103/PhysRevE.96.052118} {\bibfield  {journal} {\bibinfo  {journal} {Phys.
  Rev. E}\ }\textbf {\bibinfo {volume} {96}},\ \bibinfo {pages} {052118}
  (\bibinfo {year} {2017})}\BibitemShut {NoStop}%
\bibitem [{\citenamefont {Carollo}\ \emph {et~al.}(2018)\citenamefont
  {Carollo}, \citenamefont {Garrahan}, \citenamefont {Lesanovsky},\ and\
  \citenamefont {P\'erez-Espigares}}]{carollo18b}%
  \BibitemOpen
  \bibfield  {author} {\bibinfo {author} {\bibfnamefont {F.}~\bibnamefont
  {Carollo}}, \bibinfo {author} {\bibfnamefont {J.~P.}\ \bibnamefont
  {Garrahan}}, \bibinfo {author} {\bibfnamefont {I.}~\bibnamefont
  {Lesanovsky}}, \ and\ \bibinfo {author} {\bibfnamefont {C.}~\bibnamefont
  {P\'erez-Espigares}},\ }\bibfield  {title} {\enquote {\bibinfo {title}
  {Making rare events typical in {M}arkovian open quantum systems},}\ }\href
  {\doibase 10.1103/PhysRevA.98.010103} {\bibfield  {journal} {\bibinfo
  {journal} {Phys. Rev. A}\ }\textbf {\bibinfo {volume} {98}},\ \bibinfo
  {pages} {010103(R)} (\bibinfo {year} {2018})}\BibitemShut {NoStop}%
\bibitem [{\citenamefont {Bertini}\ \emph {et~al.}(2015)\citenamefont
  {Bertini}, \citenamefont {Sole}, \citenamefont {Gabrielli}, \citenamefont
  {Jona-Lasinio},\ and\ \citenamefont {Landim}}]{bertini15a}%
  \BibitemOpen
  \bibfield  {author} {\bibinfo {author} {\bibfnamefont {L.}~\bibnamefont
  {Bertini}}, \bibinfo {author} {\bibfnamefont {A.~De}\ \bibnamefont {Sole}},
  \bibinfo {author} {\bibfnamefont {D.}~\bibnamefont {Gabrielli}}, \bibinfo
  {author} {\bibfnamefont {G.}~\bibnamefont {Jona-Lasinio}}, \ and\ \bibinfo
  {author} {\bibfnamefont {C.}~\bibnamefont {Landim}},\ }\bibfield  {title}
  {\enquote {\bibinfo {title} {Macroscopic fluctuation theory},}\ }\href
  {http://journals.aps.org/rmp/abstract/10.1103/RevModPhys.87.593} {\bibfield
  {journal} {\bibinfo  {journal} {Rev. Mod. Phys.}\ }\textbf {\bibinfo {volume}
  {87}},\ \bibinfo {pages} {593--636} (\bibinfo {year} {2015})}\BibitemShut
  {NoStop}%
\bibitem [{\citenamefont {Touchette}(2009)}]{touchette09a}%
  \BibitemOpen
  \bibfield  {author} {\bibinfo {author} {\bibfnamefont {H.}~\bibnamefont
  {Touchette}},\ }\bibfield  {title} {\enquote {\bibinfo {title} {The large
  deviation approach to statistical mechanics},}\ }\href
  {http://dx.doi.org/10.1016/j.physrep.2009.05.002} {\bibfield  {journal}
  {\bibinfo  {journal} {Phys. Rep.}\ }\textbf {\bibinfo {volume} {478}},\
  \bibinfo {pages} {1} (\bibinfo {year} {2009})}\BibitemShut {NoStop}%
\bibitem [{\citenamefont {Lecomte}\ \emph {et~al.}(2007)\citenamefont
  {Lecomte}, \citenamefont {Appert-Rolland},\ and\ \citenamefont {van
  Wijland}}]{lecomte07c}%
  \BibitemOpen
  \bibfield  {author} {\bibinfo {author} {\bibfnamefont {V.}~\bibnamefont
  {Lecomte}}, \bibinfo {author} {\bibfnamefont {C.}~\bibnamefont
  {Appert-Rolland}}, \ and\ \bibinfo {author} {\bibfnamefont {F.}~\bibnamefont
  {van Wijland}},\ }\bibfield  {title} {\enquote {\bibinfo {title}
  {Thermodynamic formalism for systems with {M}arkov dynamics},}\ }\href
  {\doibase 10.1007/s10955-006-9254-0} {\bibfield  {journal} {\bibinfo
  {journal} {J. Stat. Phys.}\ }\textbf {\bibinfo {volume} {127}},\ \bibinfo
  {pages} {51--106} (\bibinfo {year} {2007})}\BibitemShut {NoStop}%
\bibitem [{\citenamefont {Callen}(1985)}]{callen1985}%
  \BibitemOpen
  \bibfield  {author} {\bibinfo {author} {\bibfnamefont {H.~B.}\ \bibnamefont
  {Callen}},\ }\href@noop {} {\emph {\bibinfo {title} {Thermodynamics and an
  {I}ntroduction to {T}hermostatistics}}}\ (\bibinfo  {publisher} {John Wiley
  \& Sons, 2nd Ed.},\ \bibinfo {address} {New York},\ \bibinfo {year}
  {1985})\BibitemShut {NoStop}%
\bibitem [{\citenamefont {Simon}(2009)}]{simon2009}%
  \BibitemOpen
  \bibfield  {author} {\bibinfo {author} {\bibfnamefont {D.}~\bibnamefont
  {Simon}},\ }\bibfield  {title} {\enquote {\bibinfo {title} {Construction of a
  coordinate {B}ethe ansatz for the asymmetric simple exclusion process with
  open boundaries},}\ }\href {https://doi.org/10.1088/1742-5468/2009/07/p07017}
  {\bibfield  {journal} {\bibinfo  {journal} {J. Stat. Mech. P07017}\ }
  (\bibinfo {year} {2009})}\BibitemShut {NoStop}%
\bibitem [{\citenamefont {Popkov}\ \emph {et~al.}(2010)\citenamefont {Popkov},
  \citenamefont {Sch{\"u}tz},\ and\ \citenamefont {Simon}}]{popkov10a}%
  \BibitemOpen
  \bibfield  {author} {\bibinfo {author} {\bibfnamefont {V.}~\bibnamefont
  {Popkov}}, \bibinfo {author} {\bibfnamefont {G.~M.}\ \bibnamefont
  {Sch{\"u}tz}}, \ and\ \bibinfo {author} {\bibfnamefont {D.}~\bibnamefont
  {Simon}},\ }\bibfield  {title} {\enquote {\bibinfo {title} {{ASEP} on a ring
  conditioned on enhanced flux},}\ }\href
  {http://dx.doi.org/10.1088/1742-5468/2010/10/P10007} {\bibfield  {journal}
  {\bibinfo  {journal} {J. Stat. Mech. P10007}\ } (\bibinfo {year}
  {2010})}\BibitemShut {NoStop}%
\bibitem [{\citenamefont {Jack}\ and\ \citenamefont
  {Sollich}(2010)}]{jack2010}%
  \BibitemOpen
  \bibfield  {author} {\bibinfo {author} {\bibfnamefont {R.~L.}\ \bibnamefont
  {Jack}}\ and\ \bibinfo {author} {\bibfnamefont {P.}~\bibnamefont {Sollich}},\
  }\bibfield  {title} {\enquote {\bibinfo {title} {Large deviations and
  ensembles of trajectories in stochastic models},}\ }\href
  {https://doi.org/10.1143/PTPS.184.304} {\bibfield  {journal} {\bibinfo
  {journal} {Prog. Theor. Phys. Supp.}\ }\textbf {\bibinfo {volume} {184}},\
  \bibinfo {pages} {304--317} (\bibinfo {year} {2010})}\BibitemShut {NoStop}%
\bibitem [{\citenamefont {Chetrite}\ and\ \citenamefont
  {Touchette}(2013)}]{chetrite13a}%
  \BibitemOpen
  \bibfield  {author} {\bibinfo {author} {\bibfnamefont {R.}~\bibnamefont
  {Chetrite}}\ and\ \bibinfo {author} {\bibfnamefont {H.}~\bibnamefont
  {Touchette}},\ }\bibfield  {title} {\enquote {\bibinfo {title}
  {Nonequilibrium microcanonical and canonical ensembles and their
  equivalence},}\ }\href {\doibase 10.1103/PhysRevLett.111.120601} {\bibfield
  {journal} {\bibinfo  {journal} {Phys. Rev. Lett.}\ }\textbf {\bibinfo
  {volume} {111}},\ \bibinfo {pages} {120601} (\bibinfo {year}
  {2013})}\BibitemShut {NoStop}%
\bibitem [{\citenamefont {Chetrite}\ and\ \citenamefont
  {Touchette}(2015)}]{chetrite15b}%
  \BibitemOpen
  \bibfield  {author} {\bibinfo {author} {\bibfnamefont {R.}~\bibnamefont
  {Chetrite}}\ and\ \bibinfo {author} {\bibfnamefont {H.}~\bibnamefont
  {Touchette}},\ }\bibfield  {title} {\enquote {\bibinfo {title}
  {Nonequilibrium {Markov} processes conditioned on large deviations},}\ }\href
  {https://link.springer.com/article/10.1007%2Fs00023-014-0375-8} {\bibfield
  {journal} {\bibinfo  {journal} {Ann. Henri Poincare}\ }\textbf {\bibinfo
  {volume} {16}},\ \bibinfo {pages} {2005} (\bibinfo {year}
  {2015})}\BibitemShut {NoStop}%
\bibitem [{\citenamefont {Bertini}\ \emph {et~al.}(2005)\citenamefont
  {Bertini}, \citenamefont {Sole}, \citenamefont {Gabrielli}, \citenamefont
  {Jona-Lasinio},\ and\ \citenamefont {Landim}}]{bertini05a}%
  \BibitemOpen
  \bibfield  {author} {\bibinfo {author} {\bibfnamefont {L.}~\bibnamefont
  {Bertini}}, \bibinfo {author} {\bibfnamefont {A.~De}\ \bibnamefont {Sole}},
  \bibinfo {author} {\bibfnamefont {D.}~\bibnamefont {Gabrielli}}, \bibinfo
  {author} {\bibfnamefont {G.}~\bibnamefont {Jona-Lasinio}}, \ and\ \bibinfo
  {author} {\bibfnamefont {C.}~\bibnamefont {Landim}},\ }\bibfield  {title}
  {\enquote {\bibinfo {title} {Current fluctuations in stochastic lattice
  gases},}\ }\href
  {http://journals.aps.org/prl/abstract/10.1103/PhysRevLett.94.030601}
  {\bibfield  {journal} {\bibinfo  {journal} {Phys. Rev. Lett.}\ }\textbf
  {\bibinfo {volume} {94}},\ \bibinfo {pages} {030601} (\bibinfo {year}
  {2005})}\BibitemShut {NoStop}%
\bibitem [{\citenamefont {Bodineau}\ and\ \citenamefont
  {Derrida}(2005)}]{bodineau05a}%
  \BibitemOpen
  \bibfield  {author} {\bibinfo {author} {\bibfnamefont {T.}~\bibnamefont
  {Bodineau}}\ and\ \bibinfo {author} {\bibfnamefont {B.}~\bibnamefont
  {Derrida}},\ }\bibfield  {title} {\enquote {\bibinfo {title} {Distribution of
  current in nonequilibrium diffusive systems and phase transitions},}\ }\href
  {http://journals.aps.org/pre/abstract/10.1103/PhysRevE.72.066110} {\bibfield
  {journal} {\bibinfo  {journal} {Phys. Rev. E}\ }\textbf {\bibinfo {volume}
  {72}},\ \bibinfo {pages} {066110} (\bibinfo {year} {2005})}\BibitemShut
  {NoStop}%
\bibitem [{\citenamefont {Bertini}\ \emph {et~al.}(2006)\citenamefont
  {Bertini}, \citenamefont {Sole}, \citenamefont {Gabrielli}, \citenamefont
  {Jona-Lasinio},\ and\ \citenamefont {Landim}}]{bertini06a}%
  \BibitemOpen
  \bibfield  {author} {\bibinfo {author} {\bibfnamefont {L.}~\bibnamefont
  {Bertini}}, \bibinfo {author} {\bibfnamefont {A.~De}\ \bibnamefont {Sole}},
  \bibinfo {author} {\bibfnamefont {D.}~\bibnamefont {Gabrielli}}, \bibinfo
  {author} {\bibfnamefont {G.}~\bibnamefont {Jona-Lasinio}}, \ and\ \bibinfo
  {author} {\bibfnamefont {C.}~\bibnamefont {Landim}},\ }\bibfield  {title}
  {\enquote {\bibinfo {title} {Nonequilibrium current fluctuations in
  stochastic lattice gases},}\ }\href
  {http://link.springer.com/article/10.1007/s10955-006-9056-4} {\bibfield
  {journal} {\bibinfo  {journal} {J. Stat. Phys.}\ }\textbf {\bibinfo {volume}
  {123}},\ \bibinfo {pages} {237--276} (\bibinfo {year} {2006})}\BibitemShut
  {NoStop}%
\bibitem [{\citenamefont {Hurtado}\ and\ \citenamefont
  {Garrido}(2011)}]{hurtado11a}%
  \BibitemOpen
  \bibfield  {author} {\bibinfo {author} {\bibfnamefont {P.~I.}\ \bibnamefont
  {Hurtado}}\ and\ \bibinfo {author} {\bibfnamefont {P.~L.}\ \bibnamefont
  {Garrido}},\ }\bibfield  {title} {\enquote {\bibinfo {title} {Spontaneous
  symmetry breaking at the fluctuating level},}\ }\href
  {http://journals.aps.org/prl/abstract/10.1103/PhysRevLett.107.180601}
  {\bibfield  {journal} {\bibinfo  {journal} {Phys. Rev. Lett.}\ }\textbf
  {\bibinfo {volume} {107}},\ \bibinfo {pages} {180601} (\bibinfo {year}
  {2011})}\BibitemShut {NoStop}%
\bibitem [{\citenamefont {P\'erez-Espigares}\ \emph {et~al.}(2013)\citenamefont
  {P\'erez-Espigares}, \citenamefont {Garrido},\ and\ \citenamefont
  {Hurtado}}]{perez-espigares13a}%
  \BibitemOpen
  \bibfield  {author} {\bibinfo {author} {\bibfnamefont {C. P.}~\bibnamefont
  {Espigares}}, \bibinfo {author} {\bibfnamefont {P.~L.}\ \bibnamefont
  {Garrido}}, \ and\ \bibinfo {author} {\bibfnamefont {P.~I.}\ \bibnamefont
  {Hurtado}},\ }\bibfield  {title} {\enquote {\bibinfo {title} {Dynamical phase
  transition for current statistics in a simple driven diffusive system},}\
  }\href {http://journals.aps.org/pre/abstract/10.1103/PhysRevE.87.032115}
  {\bibfield  {journal} {\bibinfo  {journal} {Phys. Rev. E}\ }\textbf {\bibinfo
  {volume} {87}},\ \bibinfo {pages} {032115} (\bibinfo {year}
  {2013})}\BibitemShut {NoStop}%
\bibitem [{\citenamefont {Jack}\ \emph {et~al.}(2015)\citenamefont {Jack},
  \citenamefont {Thompson},\ and\ \citenamefont {Sollich}}]{jack15a}%
  \BibitemOpen
  \bibfield  {author} {\bibinfo {author} {\bibfnamefont {R.~L.}\ \bibnamefont
  {Jack}}, \bibinfo {author} {\bibfnamefont {I.~R.}\ \bibnamefont {Thompson}},
  \ and\ \bibinfo {author} {\bibfnamefont {P.}~\bibnamefont {Sollich}},\
  }\bibfield  {title} {\enquote {\bibinfo {title} {Hyperuniformity and phase
  separation in biased ensembles of trajectories for diffusive systems},}\
  }\href {http://journals.aps.org/prl/abstract/10.1103/PhysRevLett.114.060601}
  {\bibfield  {journal} {\bibinfo  {journal} {Phys. Rev. Lett.}\ }\textbf
  {\bibinfo {volume} {114}},\ \bibinfo {pages} {060601} (\bibinfo {year}
  {2015})}\BibitemShut {NoStop}%
\bibitem [{\citenamefont {Nyawo}\ and\ \citenamefont
  {Touchette}(2016)}]{tsobgni16a}%
  \BibitemOpen
  \bibfield  {author} {\bibinfo {author} {\bibfnamefont {O.~Tsobgni}\
  \bibnamefont {Nyawo}}\ and\ \bibinfo {author} {\bibfnamefont
  {H.}~\bibnamefont {Touchette}},\ }\bibfield  {title} {\enquote {\bibinfo
  {title} {A minimal model of dynamical phase transition},}\ }\href
  {http://stacks.iop.org/0295-5075/116/i=5/a=50009} {\bibfield  {journal}
  {\bibinfo  {journal} {EPL}\ }\textbf {\bibinfo {volume} {116}},\
  \bibinfo {pages} {50009} (\bibinfo {year} {2016})}\BibitemShut {NoStop}%
\bibitem [{\citenamefont {Baek}\ \emph {et~al.}(2017)\citenamefont {Baek},
  \citenamefont {Kafri},\ and\ \citenamefont {Lecomte}}]{baek17a}%
  \BibitemOpen
  \bibfield  {author} {\bibinfo {author} {\bibfnamefont {Y.}~\bibnamefont
  {Baek}}, \bibinfo {author} {\bibfnamefont {Y.}~\bibnamefont {Kafri}}, \ and\
  \bibinfo {author} {\bibfnamefont {V.}~\bibnamefont {Lecomte}},\ }\bibfield
  {title} {\enquote {\bibinfo {title} {Dynamical symmetry breaking and phase
  transitions in driven diffusive systems},}\ }\href
  {http://journals.aps.org/prl/abstract/10.1103/PhysRevLett.118.030604}
  {\bibfield  {journal} {\bibinfo  {journal} {Phys. Rev. Lett.}\ }\textbf
  {\bibinfo {volume} {118}},\ \bibinfo {pages} {030604} (\bibinfo {year}
  {2017})}\BibitemShut {NoStop}%
\bibitem [{\citenamefont {Harris}\ and\ \citenamefont
  {Touchette}(2017)}]{harris17a}%
  \BibitemOpen
  \bibfield  {author} {\bibinfo {author} {\bibfnamefont {R.~J.}\ \bibnamefont
  {Harris}}\ and\ \bibinfo {author} {\bibfnamefont {H.}~\bibnamefont
  {Touchette}},\ }\bibfield  {title} {\enquote {\bibinfo {title} {Phase
  transitions in large deviations of reset processes},}\ }\href
  {http://stacks.iop.org/1751-8121/50/i=10/a=10LT01} {\bibfield  {journal}
  {\bibinfo  {journal} {J. Phys. A}\ }\textbf {\bibinfo {volume} {50}},\
  \bibinfo {pages} {10LT01} (\bibinfo {year} {2017})}\BibitemShut {NoStop}%
\bibitem [{\citenamefont {Lazarescu}(2017)}]{lazarescu17a}%
  \BibitemOpen
  \bibfield  {author} {\bibinfo {author} {\bibfnamefont {A.}~\bibnamefont
  {Lazarescu}},\ }\bibfield  {title} {\enquote {\bibinfo {title} {Generic
  dynamical phase transition in one-dimensional bulk-driven lattice gases with
  exclusion},}\ }\href {http://stacks.iop.org/1751-8121/50/i=25/a=254004}
  {\bibfield  {journal} {\bibinfo  {journal} {J. Phys. A}\ }\textbf {\bibinfo
  {volume} {50}},\ \bibinfo {pages} {254004} (\bibinfo {year}
  {2017})}\BibitemShut {NoStop}%
\bibitem [{\citenamefont {P\'erez-Espigares}\ \emph
  {et~al.}(2018{\natexlab{a}})\citenamefont {P\'erez-Espigares}, \citenamefont
  {Carollo}, \citenamefont {Garrahan},\ and\ \citenamefont
  {Hurtado}}]{perez-espigares18a}%
  \BibitemOpen
  \bibfield  {author} {\bibinfo {author} {\bibfnamefont {C.}~\bibnamefont
  {P\'erez-Espigares}}, \bibinfo {author} {\bibfnamefont {F.}~\bibnamefont
  {Carollo}}, \bibinfo {author} {\bibfnamefont {J.~P.}\ \bibnamefont
  {Garrahan}}, \ and\ \bibinfo {author} {\bibfnamefont {P.~I.}\ \bibnamefont
  {Hurtado}},\ }\bibfield  {title} {\enquote {\bibinfo {title} {Dynamical
  criticality in open systems: Nonperturbative physics, microscopic origin, and
  direct observation},}\ }\href {\doibase 10.1103/PhysRevE.98.060102}
  {\bibfield  {journal} {\bibinfo  {journal} {Phys. Rev. E}\ }\textbf {\bibinfo
  {volume} {98}},\ \bibinfo {pages} {060102(R)} (\bibinfo {year}
  {2018}{\natexlab{a}})}\BibitemShut {NoStop}%
\bibitem [{\citenamefont {P\'erez-Espigares}\ \emph
  {et~al.}(2018{\natexlab{b}})\citenamefont {P\'erez-Espigares}, \citenamefont
  {Lesanovsky}, \citenamefont {Garrahan},\ and\ \citenamefont
  {Guti\'errez}}]{perez-espigares18b}%
  \BibitemOpen
  \bibfield  {author} {\bibinfo {author} {\bibfnamefont {C.}~\bibnamefont
  {P\'erez-Espigares}}, \bibinfo {author} {\bibfnamefont {I.}~\bibnamefont
  {Lesanovsky}}, \bibinfo {author} {\bibfnamefont {J.~P.}\ \bibnamefont
  {Garrahan}}, \ and\ \bibinfo {author} {\bibfnamefont {R.}~\bibnamefont
  {Guti\'errez}},\ }\bibfield  {title} {\enquote {\bibinfo {title} {Glassy
  dynamics due to a trajectory phase transition in dissipative {R}ydberg
  gases},}\ }\href {\doibase 10.1103/PhysRevA.98.021804} {\bibfield  {journal}
  {\bibinfo  {journal} {Phys. Rev. A}\ }\textbf {\bibinfo {volume} {98}},\
  \bibinfo {pages} {021804(R)} (\bibinfo {year} {2018}{\natexlab{b}})}\BibitemShut
  {NoStop}%
\bibitem [{\citenamefont {P{\'e}rez-Espigares}\ and\ \citenamefont
  {Hurtado}(2019)}]{perez-espigares19a}%
  \BibitemOpen
  \bibfield  {author} {\bibinfo {author} {\bibfnamefont {C.}~\bibnamefont
  {P{\'e}rez-Espigares}}\ and\ \bibinfo {author} {\bibfnamefont {P.~I.}\
  \bibnamefont {Hurtado}},\ }\bibfield  {title} {\enquote {\bibinfo {title}
  {Sampling rare events across dynamical phase transitions},}\ }\href {\doibase
  10.1063/1.5091669} {\bibfield  {journal} {\bibinfo  {journal} {Chaos}\
  }\textbf {\bibinfo {volume} {29}},\ \bibinfo {pages} {083106} (\bibinfo
  {year} {2019})}\BibitemShut {NoStop}%
\bibitem [{\citenamefont {Ba\~nuls}\ and\ \citenamefont
  {Garrahan}(2019)}]{banuls19a}%
  \BibitemOpen
  \bibfield  {author} {\bibinfo {author} {\bibfnamefont {M.~C.}\ \bibnamefont
  {Ba\~nuls}}\ and\ \bibinfo {author} {\bibfnamefont {J.~P.}\ \bibnamefont
  {Garrahan}},\ }\bibfield  {title} {\enquote {\bibinfo {title} {Using matrix
  product states to study the dynamical large deviations of kinetically
  constrained models},}\ }\href {\doibase 10.1103/PhysRevLett.123.200601}
  {\bibfield  {journal} {\bibinfo  {journal} {Phys. Rev. Lett.}\ }\textbf
  {\bibinfo {volume} {123}},\ \bibinfo {pages} {200601} (\bibinfo {year}
  {2019})}\BibitemShut {NoStop}%
\bibitem [{\citenamefont {Hurtado-Guti\'errez}\ \emph
  {et~al.}(2020)\citenamefont {Hurtado-Guti\'errez}, \citenamefont {Carollo},
  \citenamefont {P\'erez-Espigares},\ and\ \citenamefont
  {Hurtado}}]{hurtado-guti20a}%
  \BibitemOpen
  \bibfield  {author} {\bibinfo {author} {\bibfnamefont {R.}~\bibnamefont
  {Hurtado-Guti\'errez}}, \bibinfo {author} {\bibfnamefont {F.}~\bibnamefont
  {Carollo}}, \bibinfo {author} {\bibfnamefont {C.}~\bibnamefont
  {P\'erez-Espigares}}, \ and\ \bibinfo {author} {\bibfnamefont {P.~I.}\
  \bibnamefont {Hurtado}},\ }\bibfield  {title} {\enquote {\bibinfo {title}
  {Building continuous time crystals from rare events},}\ }\href {\doibase
  10.1103/PhysRevLett.125.160601} {\bibfield  {journal} {\bibinfo  {journal}
  {Phys. Rev. Lett.}\ }\textbf {\bibinfo {volume} {125}},\ \bibinfo {pages}
  {160601} (\bibinfo {year} {2020})}\BibitemShut {NoStop}%
\bibitem [{\citenamefont {Hurtado}\ \emph {et~al.}(2011)\citenamefont
  {Hurtado}, \citenamefont {P\'erez-Espigares}, \citenamefont {del Pozo},\ and\
  \citenamefont {Garrido}}]{hurtado11b}%
  \BibitemOpen
  \bibfield  {author} {\bibinfo {author} {\bibfnamefont {P.~I.}\ \bibnamefont
  {Hurtado}}, \bibinfo {author} {\bibfnamefont {C.}~\bibnamefont
  {P\'erez-Espigares}}, \bibinfo {author} {\bibfnamefont {J.~J.}\ \bibnamefont
  {del Pozo}}, \ and\ \bibinfo {author} {\bibfnamefont {P.~L.}\ \bibnamefont
  {Garrido}},\ }\bibfield  {title} {\enquote {\bibinfo {title} {Symmetries in
  fluctuations far from equilibrium},}\ }\href
  {http://www.pnas.org/content/108/19/7704.short} {\bibfield  {journal}
  {\bibinfo  {journal} {Proc. Natl. Acad. Sci. USA}\ }\textbf {\bibinfo
  {volume} {108}},\ \bibinfo {pages} {7704--7709} (\bibinfo {year}
  {2011})}\BibitemShut {NoStop}%
\bibitem [{\citenamefont {Villavicencio-Sanchez}\ \emph
  {et~al.}(2014)\citenamefont {Villavicencio-Sanchez}, \citenamefont {Harris},\
  and\ \citenamefont {Touchette}}]{villavicencio14a}%
  \BibitemOpen
  \bibfield  {author} {\bibinfo {author} {\bibfnamefont {R.}~\bibnamefont
  {Villavicencio-Sanchez}}, \bibinfo {author} {\bibfnamefont {R.~J.}\
  \bibnamefont {Harris}}, \ and\ \bibinfo {author} {\bibfnamefont
  {H.}~\bibnamefont {Touchette}},\ }\bibfield  {title} {\enquote {\bibinfo
  {title} {Fluctuation relations for anisotropic systems},}\ }\href
  {http://iopscience.iop.org/article/10.1209/0295-5075/105/30009/meta}
  {\bibfield  {journal} {\bibinfo  {journal} {EPL}\ }\textbf
  {\bibinfo {volume} {105}},\ \bibinfo {pages} {30009} (\bibinfo {year}
  {2014})}\BibitemShut {NoStop}%
\bibitem [{\citenamefont {P\'erez-Espigares}\ \emph {et~al.}(2016)\citenamefont
  {P\'erez-Espigares}, \citenamefont {Garrido},\ and\ \citenamefont
  {Hurtado}}]{perez-espigares16a}%
  \BibitemOpen
  \bibfield  {author} {\bibinfo {author} {\bibfnamefont {C.}~\bibnamefont
  {P\'erez-Espigares}}, \bibinfo {author} {\bibfnamefont {P.~L.}\ \bibnamefont
  {Garrido}}, \ and\ \bibinfo {author} {\bibfnamefont {P.~I.}\ \bibnamefont
  {Hurtado}},\ }\bibfield  {title} {\enquote {\bibinfo {title} {Weak additivity
  principle for current statistics in $d$-dimensions},}\ }\href
  {http://journals.aps.org/pre/abstract/10.1103/PhysRevE.93.040103} {\bibfield
  {journal} {\bibinfo  {journal} {Phys. Rev. E}\ }\textbf {\bibinfo {volume}
  {93}},\ \bibinfo {pages} {040103(R)} (\bibinfo {year} {2016})}\BibitemShut
  {NoStop}%
\bibitem [{\citenamefont {Tiz{\'o}n-Escamilla}\ \emph
  {et~al.}(2017)\citenamefont {Tiz{\'o}n-Escamilla}, \citenamefont
  {P{\'e}rez-Espigares}, \citenamefont {Garrido},\ and\ \citenamefont
  {Hurtado}}]{tizon-escamilla17b}%
  \BibitemOpen
  \bibfield  {author} {\bibinfo {author} {\bibfnamefont {N.}~\bibnamefont
  {Tiz{\'o}n-Escamilla}}, \bibinfo {author} {\bibfnamefont {C.}~\bibnamefont
  {P{\'e}rez-Espigares}}, \bibinfo {author} {\bibfnamefont {P.~L.}\
  \bibnamefont {Garrido}}, \ and\ \bibinfo {author} {\bibfnamefont {P.~I.}\
  \bibnamefont {Hurtado}},\ }\bibfield  {title} {\enquote {\bibinfo {title}
  {Order and symmetry-breaking in the fluctuations of driven systems},}\ }\href
  {\doibase 10.1103/PhysRevLett.119.090602} {\bibfield  {journal} {\bibinfo
  {journal} {Phys. Rev. Lett.}\ }\textbf {\bibinfo {volume} {119}},\ \bibinfo
  {pages} {090602} (\bibinfo {year} {2017})}\BibitemShut {NoStop}%
\bibitem [{\citenamefont {Jack}\ \emph {et~al.}(2020)\citenamefont {Jack},
  \citenamefont {Nemoto},\ and\ \citenamefont {Lecomte}}]{jack20a}%
  \BibitemOpen
  \bibfield  {author} {\bibinfo {author} {\bibfnamefont {R.~L.}\ \bibnamefont
  {Jack}}, \bibinfo {author} {\bibfnamefont {T.}~\bibnamefont {Nemoto}}, \ and\
  \bibinfo {author} {\bibfnamefont {V.}~\bibnamefont {Lecomte}},\ }\bibfield
  {title} {\enquote {\bibinfo {title} {Dynamical phase coexistence in the
  {F}redrickson{\textendash}{A}ndersen model},}\ }\href {\doibase
  10.1088/1742-5468/ab7af6} {\bibfield  {journal} {\bibinfo  {journal} {J.
  Stat. Mech.}\ }\textbf {\bibinfo {volume} {2020}},\ \bibinfo {pages} {053204}
  (\bibinfo {year} {2020})}\BibitemShut {NoStop}%
\bibitem [{\citenamefont {Casert}\ \emph {et~al.}(2020)\citenamefont {Casert},
  \citenamefont {Vieijra}, \citenamefont {Whitelam},\ and\ \citenamefont
  {Tamblyn}}]{casert2020}%
  \BibitemOpen
  \bibfield  {author} {\bibinfo {author} {\bibfnamefont {C.}~\bibnamefont
  {Casert}}, \bibinfo {author} {\bibfnamefont {T.}~\bibnamefont {Vieijra}},
  \bibinfo {author} {\bibfnamefont {S.}~\bibnamefont {Whitelam}}, \ and\
  \bibinfo {author} {\bibfnamefont {I.}~\bibnamefont {Tamblyn}},\ }\bibfield
  {title} {\enquote {\bibinfo {title} {Dynamical large deviations of
  two-dimensional kinetically constrained models using a neural-network state
  ansatz},}\ }\href@noop {} {\bibfield  {journal} {\bibinfo  {journal}
  {arXiv:2011.08657}\ } (\bibinfo {year} {2020})}\BibitemShut {NoStop}%
\bibitem [{\citenamefont {Bodineau}\ \emph {et~al.}(2008)\citenamefont
  {Bodineau}, \citenamefont {Derrida},\ and\ \citenamefont
  {Lebowitz}}]{bodineau08a}%
  \BibitemOpen
  \bibfield  {author} {\bibinfo {author} {\bibfnamefont {T.}~\bibnamefont
  {Bodineau}}, \bibinfo {author} {\bibfnamefont {B.}~\bibnamefont {Derrida}}, \
  and\ \bibinfo {author} {\bibfnamefont {J.L.}\ \bibnamefont {Lebowitz}},\
  }\bibfield  {title} {\enquote {\bibinfo {title} {Vortices in the
  two-dimensional simple exclusion process},}\ }\href
  {http://link.springer.com/article/10.1007/s10955-008-9518-y} {\bibfield
  {journal} {\bibinfo  {journal} {J. Stat. Phys.}\ }\textbf {\bibinfo {volume}
  {131}},\ \bibinfo {pages} {821} (\bibinfo {year} {2008})}\BibitemShut
  {NoStop}%
\bibitem [{\citenamefont {Kumar}\ \emph {et~al.}(2015)\citenamefont {Kumar},
  \citenamefont {Soni}, \citenamefont {Ramaswamy},\ and\ \citenamefont
  {Sood}}]{kumar15a}%
  \BibitemOpen
  \bibfield  {author} {\bibinfo {author} {\bibfnamefont {N.}~\bibnamefont
  {Kumar}}, \bibinfo {author} {\bibfnamefont {H.}~\bibnamefont {Soni}},
  \bibinfo {author} {\bibfnamefont {S.}~\bibnamefont {Ramaswamy}}, \ and\
  \bibinfo {author} {\bibfnamefont {A.~K.}\ \bibnamefont {Sood}},\ }\bibfield
  {title} {\enquote {\bibinfo {title} {Anisotropic isometric fluctuation
  relations in experiment and theory on a self-propelled rod},}\ }\href
  {http://journals.aps.org/pre/abstract/10.1103/PhysRevE.91.030102} {\bibfield
  {journal} {\bibinfo  {journal} {Phys. Rev. E}\ }\textbf {\bibinfo {volume}
  {91}},\ \bibinfo {pages} {030102(R)} (\bibinfo {year} {2015})}\BibitemShut
  {NoStop}%
\bibitem [{\citenamefont {Falasco}\ \emph {et~al.}(2016)\citenamefont
  {Falasco}, \citenamefont {Pfaller}, \citenamefont {Bregulla}, \citenamefont
  {Cichos},\ and\ \citenamefont {Kroy}}]{falasco16a}%
  \BibitemOpen
  \bibfield  {author} {\bibinfo {author} {\bibfnamefont {G.}~\bibnamefont
  {Falasco}}, \bibinfo {author} {\bibfnamefont {R.}~\bibnamefont {Pfaller}},
  \bibinfo {author} {\bibfnamefont {A.~P.}\ \bibnamefont {Bregulla}}, \bibinfo
  {author} {\bibfnamefont {F.}~\bibnamefont {Cichos}}, \ and\ \bibinfo {author}
  {\bibfnamefont {K.}~\bibnamefont {Kroy}},\ }\bibfield  {title} {\enquote
  {\bibinfo {title} {Exact symmetries in the velocity fluctuations of a hot
  {B}rownian swimmer},}\ }\href {\doibase 10.1103/PhysRevE.94.030602}
  {\bibfield  {journal} {\bibinfo  {journal} {Phys. Rev. E}\ }\textbf {\bibinfo
  {volume} {94}},\ \bibinfo {pages} {030602(R)} (\bibinfo {year}
  {2016})}\BibitemShut {NoStop}%
\bibitem [{\citenamefont {Ciliberto}\ and\ \citenamefont
  {Laroche}(1998)}]{ciliberto98a}%
  \BibitemOpen
  \bibfield  {author} {\bibinfo {author} {\bibfnamefont {S.}~\bibnamefont
  {Ciliberto}}\ and\ \bibinfo {author} {\bibfnamefont {C.}~\bibnamefont
  {Laroche}},\ }\bibfield  {title} {\enquote {\bibinfo {title} {An experimental
  test of the {G}allavotti-{C}ohen fluctuation theorem},}\ }\href@noop {}
  {\bibfield  {journal} {\bibinfo  {journal} {J. Phys. IV} France\ }\textbf {\bibinfo
  {volume} {8}},\ \bibinfo {pages} {215} (\bibinfo {year} {1998})}\BibitemShut
  {NoStop}%
\bibitem [{\citenamefont {Feitosa}\ and\ \citenamefont
  {Menon}(2004)}]{feitosa04a}%
  \BibitemOpen
  \bibfield  {author} {\bibinfo {author} {\bibfnamefont {K.}~\bibnamefont
  {Feitosa}}\ and\ \bibinfo {author} {\bibfnamefont {N.}~\bibnamefont
  {Menon}},\ }\bibfield  {title} {\enquote {\bibinfo {title} {Fluidized
  granular medium as an instance of the fluctuation theorem},}\ }\href
  {\doibase 10.1103/PhysRevLett.92.164301} {\bibfield  {journal} {\bibinfo
  {journal} {Phys. Rev. Lett.}\ }\textbf {\bibinfo {volume} {92}},\ \bibinfo
  {pages} {164301} (\bibinfo {year} {2004})}\BibitemShut {NoStop}%
\bibitem [{Note1()}]{Note1}%
  \BibitemOpen
  \bibinfo {note} {The normalization of the Doob-generator eigenvectors is such
  that the largest absolute value of the components of ${\protect \bf
  l}^D_{s,i}$ is one, and $({\protect \bf l}^D_{s,i})^T {\protect \bf
  r}^D_{s,i} = 1$. Therefore, ${\protect \bf l}^D_{s,0} = (1,1,\protect \ldots
  ,1)^T$ and the sum of all the components of ${\protect \bf r}_{s,0}^D$ is one
  (${\protect \bf p}^s_{\protect \text {stat}} = {\protect \bf r}_{s,0}^D$ is a
  probability vector).}\BibitemShut {Stop}%
\bibitem [{\citenamefont {Guti\'errez}\ and\ \citenamefont
  {P\'erez-Espigares}(2021)}]{gutierrez2021}%
  \BibitemOpen
  \bibfield  {author} {\bibinfo {author} {\bibfnamefont {R.}~\bibnamefont
  {Guti\'errez}}\ and\ \bibinfo {author} {\bibfnamefont {C.}~\bibnamefont
  {P\'erez-Espigares}},\ }\bibfield  {title} {\enquote {\bibinfo {title}
  {Generalized optimal paths and weight distributions revealed through the
  large deviations of random walks on networks},}\ }\href {\doibase
  10.1103/PhysRevE.103.022319} {\bibfield  {journal} {\bibinfo  {journal}
  {Phys. Rev. E}\ }\textbf {\bibinfo {volume} {103}},\ \bibinfo {pages}
  {022319} (\bibinfo {year} {2021})}\BibitemShut {NoStop}%
\bibitem [{\citenamefont {Villavicencio-Sanchez}\ \emph
  {et~al.}(2012)\citenamefont {Villavicencio-Sanchez}, \citenamefont {Harris},\
  and\ \citenamefont {Touchette}}]{villavicencio12a}%
  \BibitemOpen
  \bibfield  {author} {\bibinfo {author} {\bibfnamefont {R.}~\bibnamefont
  {Villavicencio-Sanchez}}, \bibinfo {author} {\bibfnamefont {R.~J.}\
  \bibnamefont {Harris}}, \ and\ \bibinfo {author} {\bibfnamefont
  {H.}~\bibnamefont {Touchette}},\ }\bibfield  {title} {\enquote {\bibinfo
  {title} {Current loops and fluctuations in the zero-range process on a
  diamond lattice},}\ }\href@noop {} {\bibfield  {journal} {\bibinfo  {journal}
  {J. Stat. Mech. P07007}\ } (\bibinfo {year} {2012})}\BibitemShut {NoStop}%
\bibitem [{\citenamefont {Spohn}(2012)}]{spohn12a}%
  \BibitemOpen
  \bibfield  {author} {\bibinfo {author} {\bibfnamefont {H.}~\bibnamefont
  {Spohn}},\ }\href {https://www.springer.com/la/book/9783642843730} {\emph
  {\bibinfo {title} {{Large Scale Dynamics of Interacting Particles}}}},\
  {Theoretical and Mathematical Physics}\ (\bibinfo  {publisher} {Springer,
  Berlin},\ \bibinfo {year} {2012})\BibitemShut {NoStop}%
\bibitem [{\citenamefont {Gaveau}\ and\ \citenamefont
  {Schulman}(2006)}]{gaveau06a}%
  \BibitemOpen
  \bibfield  {author} {\bibinfo {author} {\bibfnamefont {B.}~\bibnamefont
  {Gaveau}}\ and\ \bibinfo {author} {\bibfnamefont {L.~S.}\ \bibnamefont
  {Schulman}},\ }\bibfield  {title} {\enquote {\bibinfo {title} {Multiple
  phases in stochastic dynamics: Geometry and probabilities},}\ }\href
  {\doibase 10.1103/PhysRevE.73.036124} {\bibfield  {journal} {\bibinfo
  {journal} {Phys. Rev. E}\ }\textbf {\bibinfo {volume} {73}},\ \bibinfo
  {pages} {036124} (\bibinfo {year} {2006})}\BibitemShut {NoStop}%
\bibitem [{\citenamefont {Minganti}\ \emph {et~al.}(2018)\citenamefont
  {Minganti}, \citenamefont {Biella}, \citenamefont {Bartolo},\ and\
  \citenamefont {Ciuti}}]{minganti18a}%
  \BibitemOpen
  \bibfield  {author} {\bibinfo {author} {\bibfnamefont {F.}~\bibnamefont
  {Minganti}}, \bibinfo {author} {\bibfnamefont {A.}~\bibnamefont {Biella}},
  \bibinfo {author} {\bibfnamefont {N.}~\bibnamefont {Bartolo}}, \ and\
  \bibinfo {author} {\bibfnamefont {C.}~\bibnamefont {Ciuti}},\ }\bibfield
  {title} {\enquote {\bibinfo {title} {Spectral theory of {L}iouvillians for
  dissipative phase transitions},}\ }\href {\doibase
  10.1103/PhysRevA.98.042118} {\bibfield  {journal} {\bibinfo  {journal} {Phys.
  Rev. A}\ }\textbf {\bibinfo {volume} {98}},\ \bibinfo {pages} {042118}
  (\bibinfo {year} {2018})}\BibitemShut {NoStop}%
\bibitem [{\citenamefont {Thouless}(1974)}]{thouless1974}%
  \BibitemOpen
  \bibfield  {author} {\bibinfo {author} {\bibfnamefont {D.~J}\ \bibnamefont
  {Thouless}},\ }\bibfield  {title} {\enquote {\bibinfo {title} {Electrons in
  disordered systems and the theory of localization},}\ }\href@noop {}
  {\bibfield  {journal} {\bibinfo  {journal} {Phys. Rep.}\ }\textbf {\bibinfo
  {volume} {13}},\ \bibinfo {pages} {93--142} (\bibinfo {year}
  {1974})}\BibitemShut {NoStop}%
\bibitem [{\citenamefont {Burda}\ \emph {et~al.}(2009)\citenamefont {Burda},
  \citenamefont {Duda}, \citenamefont {Luck},\ and\ \citenamefont
  {Waclaw}}]{burda2009}%
  \BibitemOpen
  \bibfield  {author} {\bibinfo {author} {\bibfnamefont {Z.}~\bibnamefont
  {Burda}}, \bibinfo {author} {\bibfnamefont {J.}~\bibnamefont {Duda}},
  \bibinfo {author} {\bibfnamefont {J.~M.}\ \bibnamefont {Luck}}, \ and\
  \bibinfo {author} {\bibfnamefont {B.}~\bibnamefont {Waclaw}},\ }\bibfield
  {title} {\enquote {\bibinfo {title} {Localization of the maximal entropy
  random walk},}\ }\href {\doibase 10.1103/PhysRevLett.102.160602} {\bibfield
  {journal} {\bibinfo  {journal} {Phys. Rev. Lett.}\ }\textbf {\bibinfo
  {volume} {102}},\ \bibinfo {pages} {160602} (\bibinfo {year}
  {2009})}\BibitemShut {NoStop}%
\bibitem [{\citenamefont {Coghi}\ \emph {et~al.}(2019)\citenamefont {Coghi},
  \citenamefont {Morand},\ and\ \citenamefont {Touchette}}]{coghi2019}%
  \BibitemOpen
  \bibfield  {author} {\bibinfo {author} {\bibfnamefont {F.}~\bibnamefont
  {Coghi}}, \bibinfo {author} {\bibfnamefont {J.}~\bibnamefont {Morand}}, \
  and\ \bibinfo {author} {\bibfnamefont {H.}~\bibnamefont {Touchette}},\
  }\bibfield  {title} {\enquote {\bibinfo {title} {Large deviations of random
  walks on random graphs},}\ }\href {\doibase 10.1103/PhysRevE.99.022137}
  {\bibfield  {journal} {\bibinfo  {journal} {Phys. Rev. E}\ }\textbf {\bibinfo
  {volume} {99}},\ \bibinfo {pages} {022137} (\bibinfo {year}
  {2019})}\BibitemShut {NoStop}%
\bibitem [{\citenamefont {Bacco}\ \emph {et~al.}(2016)\citenamefont {Bacco},
  \citenamefont {Guggiola}, \citenamefont {K{\"u}hn},\ and\ \citenamefont
  {Paga}}]{bacco16a}%
  \BibitemOpen
  \bibfield  {author} {\bibinfo {author} {\bibfnamefont {C.~De}\ \bibnamefont
  {Bacco}}, \bibinfo {author} {\bibfnamefont {A.}~\bibnamefont {Guggiola}},
  \bibinfo {author} {\bibfnamefont {R.}~\bibnamefont {K{\"u}hn}}, \ and\
  \bibinfo {author} {\bibfnamefont {P.}~\bibnamefont {Paga}},\ }\bibfield
  {title} {\enquote {\bibinfo {title} {Rare events statistics of random walks
  on networks: localisation and other dynamical phase transitions},}\ }\href
  {\doibase 10.1088/1751-8113/49/18/184003} {\bibfield  {journal} {\bibinfo
  {journal} {J. Phys. A}\ }\textbf {\bibinfo {volume} {49}},\ \bibinfo {pages}
  {184003} (\bibinfo {year} {2016})}\BibitemShut {NoStop}%
\bibitem [{\citenamefont {Attal}\ \emph {et~al.}(2012)\citenamefont {Attal},
  \citenamefont {Petruccione},\ and\ \citenamefont {Sinayskiy}}]{attal2012}%
  \BibitemOpen
  \bibfield  {author} {\bibinfo {author} {\bibfnamefont {S.}~\bibnamefont
  {Attal}}, \bibinfo {author} {\bibfnamefont {F.}~\bibnamefont {Petruccione}},
  \ and\ \bibinfo {author} {\bibfnamefont {I.}~\bibnamefont {Sinayskiy}},\
  }\bibfield  {title} {\enquote {\bibinfo {title} {Open quantum walks on
  graphs},}\ }\href {\doibase https://doi.org/10.1016/j.physleta.2012.03.040}
  {\bibfield  {journal} {\bibinfo  {journal} {Phys. Lett. A}\ }\textbf
  {\bibinfo {volume} {376}},\ \bibinfo {pages} {1545--1548} (\bibinfo {year}
  {2012})}\BibitemShut {NoStop}%
\bibitem [{\citenamefont {Garnerone}(2012)}]{garnerone2012}%
  \BibitemOpen
  \bibfield  {author} {\bibinfo {author} {\bibfnamefont {S.}~\bibnamefont
  {Garnerone}},\ }\bibfield  {title} {\enquote {\bibinfo {title} {Thermodynamic
  formalism for dissipative quantum walks},}\ }\href {\doibase
  10.1103/PhysRevA.86.032342} {\bibfield  {journal} {\bibinfo  {journal} {Phys.
  Rev. A}\ }\textbf {\bibinfo {volume} {86}},\ \bibinfo {pages} {032342}
  (\bibinfo {year} {2012})}\BibitemShut {NoStop}%
\bibitem [{\citenamefont {Macieszczak}\ \emph {et~al.}(2016)\citenamefont
  {Macieszczak}, \citenamefont {Gu\ifmmode \mbox{\c{t}}\else
  \c{t}\fi{}\ifmmode~\u{a}\else \u{a}\fi{}}, \citenamefont {Lesanovsky},\ and\
  \citenamefont {Garrahan}}]{macieszczak16a}%
  \BibitemOpen
  \bibfield  {author} {\bibinfo {author} {\bibfnamefont {K.}~\bibnamefont
  {Macieszczak}}, \bibinfo {author} {\bibfnamefont {M.}~\bibnamefont
  {Gu\ifmmode \mbox{\c{t}}\else \c{t}\fi{}\ifmmode~\u{a}\else \u{a}\fi{}}},
  \bibinfo {author} {\bibfnamefont {I.}~\bibnamefont {Lesanovsky}}, \ and\
  \bibinfo {author} {\bibfnamefont {J.~P.}\ \bibnamefont {Garrahan}},\
  }\bibfield  {title} {\enquote {\bibinfo {title} {Towards a theory of
  metastability in open quantum dynamics},}\ }\href {\doibase
  10.1103/PhysRevLett.116.240404} {\bibfield  {journal} {\bibinfo  {journal}
  {Phys. Rev. Lett.}\ }\textbf {\bibinfo {volume} {116}},\ \bibinfo {pages}
  {240404} (\bibinfo {year} {2016})}\BibitemShut {NoStop}%
\bibitem [{\citenamefont {Anderson}(1958)}]{anderson1958}%
  \BibitemOpen
  \bibfield  {author} {\bibinfo {author} {\bibfnamefont {P.~W.}\ \bibnamefont
  {Anderson}},\ }\bibfield  {title} {\enquote {\bibinfo {title} {Absence of
  diffusion in certain random lattices},}\ }\href {\doibase
  10.1103/PhysRev.109.1492} {\bibfield  {journal} {\bibinfo  {journal} {Phys.
  Rev.}\ }\textbf {\bibinfo {volume} {109}},\ \bibinfo {pages} {1492--1505}
  (\bibinfo {year} {1958})}\BibitemShut {NoStop}%
\bibitem [{\citenamefont {Nandkishore}\ and\ \citenamefont
  {Huse}(2015)}]{nandkishore2015}%
  \BibitemOpen
  \bibfield  {author} {\bibinfo {author} {\bibfnamefont {R.}~\bibnamefont
  {Nandkishore}}\ and\ \bibinfo {author} {\bibfnamefont {D.~A.}\ \bibnamefont
  {Huse}},\ }\bibfield  {title} {\enquote {\bibinfo {title} {Many-body
  localization and thermalization in quantum statistical mechanics},}\ }\href
  {\doibase 10.1146/annurev-conmatphys-031214-014726} {\bibfield  {journal}
  {\bibinfo  {journal} {Annu. Rev. Condens. Matter Phys.}\ }\textbf {\bibinfo
  {volume} {6}},\ \bibinfo {pages} {15--38} (\bibinfo {year}
  {2015})}\BibitemShut {NoStop}%
\bibitem [{\citenamefont {{Iannone}}\ \emph {et~al.}(2019)\citenamefont
  {{Iannone}}, \citenamefont {{Ambrosino}}, \citenamefont {{Bracco}},
  \citenamefont {{De Rosa}}, \citenamefont {{Funel}}, \citenamefont
  {{Guarnieri}}, \citenamefont {{Migliori}}, \citenamefont {{Palombi}},
  \citenamefont {{Ponti}}, \citenamefont {{Santomauro}},\ and\ \citenamefont
  {{Procacci}}}]{iannone2019}%
  \BibitemOpen
  \bibfield  {author} {\bibinfo {author} {\bibfnamefont {F.}~\bibnamefont
  {{Iannone}}}, \bibinfo {author} {\bibfnamefont {F.}~\bibnamefont
  {{Ambrosino}}}, \bibinfo {author} {\bibfnamefont {G.}~\bibnamefont
  {{Bracco}}}, \bibinfo {author} {\bibfnamefont {M.}~\bibnamefont {{De Rosa}}},
  \bibinfo {author} {\bibfnamefont {A.}~\bibnamefont {{Funel}}}, \bibinfo
  {author} {\bibfnamefont {G.}~\bibnamefont {{Guarnieri}}}, \bibinfo {author}
  {\bibfnamefont {S.}~\bibnamefont {{Migliori}}}, \bibinfo {author}
  {\bibfnamefont {F.}~\bibnamefont {{Palombi}}}, \bibinfo {author}
  {\bibfnamefont {G.}~\bibnamefont {{Ponti}}}, \bibinfo {author} {\bibfnamefont
  {G.}~\bibnamefont {{Santomauro}}}, \ and\ \bibinfo {author} {\bibfnamefont
  {P.}~\bibnamefont {{Procacci}}},\ }\bibfield  {title} {\enquote {\bibinfo
  {title} {{CRESCO ENEA HPC} clusters: a working example of a multifabric
  {GPFS} {S}pectrum {S}cale layout},}\ }in\ \href@noop {} {\emph {\bibinfo
  {booktitle} {2019 {I}nternational {C}onference on {H}igh {P}erformance
  {C}omputing {S}imulation ({HPCS})}}}\ (\bibinfo {year} {2019})\ p.\ \bibinfo
  {pages} {1051}\BibitemShut {NoStop}%
\end{thebibliography}

%

\end{document}